
\newcount\mgnf\newcount\tipi\newcount\tipoformule
\newcount\aux\newcount\driver\newcount\cind\global\newcount\bz
\newcount\tipobib\newcount\stile\newcount\modif

\newdimen\stdindent\newdimen\bibskip
\newdimen\maxit\maxit=0pt

\stile=0         
\tipobib=1       
\bz=0            
\cind=0          
\mgnf=1          
\tipoformule=0   
\aux=0           


\ifnum\mgnf=0
   \magnification=\magstep0 
   \hsize=17truecm\vsize=24truecm\hoffset=-1cm
   \parindent=4.pt\stdindent=\parindent\fi
\ifnum\mgnf=1
   \magnification=\magstep1\hoffset=-0.7truecm
   \voffset=-1.0truecm\hsize=18truecm\vsize=24.truecm
   \baselineskip=14pt plus0.1pt minus0.1pt \parindent=6pt
   \lineskip=4pt\lineskiplimit=0.1pt      \parskip=0.1pt plus1pt
   \stdindent=\parindent\fi


\def\fine#1{}
\def\draft#1{\bz=1\ifnum\mgnf=1\baselineskip=22pt 
			      \else\baselineskip=16pt\fi
   \ifnum\stile=0\headline={\hfill DRAFT #1}\fi\raggedbottom
    \setbox150\vbox{\parindent=0pt\centerline{\bf Figures' captions}\*}
    \def\gnuins ##1 ##2 ##3{\gnuinsf {##1} {##2} {##3}}
    \def\gnuin ##1 ##2 ##3 ##4 ##5 ##6{\gnuinf {##1} {##2} {##3} 
		{##4} {##5} {##6}} 
    \def\eqfig##1##2##3##4##5##6{\eqfigf {##1} {##2} {##3} {##4} {##5} {##6}}
    \def\eqfigfor##1##2##3##4##5##6##7
             {\eqfigforf {##1} {##2} {##3} {##4} {##5} {##6} {##7}}
      \def\fine ##1{\vfill\eject
	 	\def\geq(####1){}
               \unvbox150\vfill\eject\raggedbottom
                \centerline{FIGURES}\unvbox149 ##1}}


\newcount\prau

\def\titolo#1{\setbox197\vbox{ 
\leftskip=0pt plus16em \rightskip=\leftskip
\spaceskip=.3333em \xspaceskip=.5em \parfillskip=0pt
\pretolerance=9999  \tolerance=9999
\hyphenpenalty=9999 \exhyphenpenalty=9999
\ftitolo #1}}
\def\abstract#1{\setbox198\vbox{
     \centerline{\vbox{\advance\hsize by -2cm \parindent=0pt\it Abstact: #1}}}}
\def\parole#1{\setbox195\hbox{
     \centerline{\vbox{\advance\hsize by -2cm \parindent=0pt Keywords: #1.}}}}
\def\autore#1#2{\setbox199\hbox{\unhbox199\ifnum\prau=0 #1%
\else, #1\fi\global\advance\prau by 1$^{\simbau}$}
     \setbox196\vbox {\advance\hsize by -\parindent\copy196$^{\simbau}${#2}}}
\def\prima{\unvbox197\vskip1truecm\centerline{\unhbox199}
     \footnote{}{\unvbox196}\vskip1truecm\unvbox198\vskip1truecm\copy195}
\def\simbau{\ifcase\prau
	\or \dagger \or \ddagger \or * \or \star \or \bullet\fi}


      \let\l=\lambda


{\count255=\time\divide\count255 by 60 \xdef\oramin{\number\count255}
        \multiply\count255 by-60\advance\count255 by\time
   \xdef\oramin{\oramin:\ifnum\count255<10 0\fi\the\count255}}
\def\ora{\oramin }

\def\data{\number\day/\ifcase\month\or gennaio \or febbraio \or marzo \or
aprile \or maggio \or giugno \or luglio \or agosto \or settembre
\or ottobre \or novembre \or dicembre \fi/\number\year;\ \ora}

\setbox200\hbox{$\scriptscriptstyle \data $}


\newcount\pgn \pgn=1
\newcount\firstpage

\def\foglio{\number\numsec:\number\pgn\global\advance\pgn by 1}
\def\foglioa{A\number\numsec:\number\pgn\global\advance\pgn by 1}

\def\pagina{\vfill\eject}
\def\ppagina{\ifodd\pageno\pagina\null\pagina\else\pagina\fi}
\def\ppaginan{\ifodd-\pageno\pagina\null\pagina\else\pagina\fi}

\def\setind{\firstpage=\pageno}
\def\setcap#1{\null\def\titlecap{#1}\global\firstpage=\pageno}
\def\titletesi{Indici critici per sistemi fermionici in una dimensione}

\ifnum\stile=1
  \def\pagenumbers{\headline={%
  \ifnum\pageno=\firstpage\hfil\else%
     \ifodd\pageno\hfill{\sc\titlecap}~~{\bf\folio}%
      \else{\bf\folio}~~{\sc\titletesi}\hfill\fi\fi}
  \footline={\ifnum\bz=0
                   \hfill\else\rlap{\hbox{\copy200}\ $\st[\foglio]$}\hfill\fi}}
  \def\pagenumbersind{\headline={%
  \ifnum\pageno=\firstpage\hfil\else%
    \ifodd\pageno\hfill{\rm\romannumeral-\pageno}%
     \else{\rm\romannumeral-\pageno}\hfill\fi\fi}
  \footline={\ifnum\bz=0
                   \hfill\else\rlap{\hbox{\copy200}\ $\st[\foglio]$}\hfill\fi}}
\else
  \def\pagenumbers{\headline={\hfill}
     \footline={\ifnum\bz=0\hfill\folio\hfill
                \else\rlap{\hbox{\copy200}\ $\st[\foglio]$}
		   \hfill\folio\hfill\fi}}
\fi

\pagenumbers

\def\numeropag#1{
   \ifnum #1<0 \romannumeral -#1\else \number #1\fi
   }


\global\newcount\numsec\global\newcount\numfor
\global\newcount\numfig\global\newcount\numpar
\global\newcount\numteo\global\newcount\numlem

\numfig=1\numsec=0

\gdef\profonditastruttura{\dp\strutbox}
\def\senondefinito#1{\expandafter\ifx\csname#1\endcsname\relax}
\def\SIA #1,#2,#3 {\senondefinito{#1#2}%
\expandafter\xdef\csname#1#2\endcsname{#3}\else%
\write16{???? ma #1,#2 e' gia' stato definito !!!!}\fi}
\def\etichetta(#1){(\veroparagrafo.\veraformula)
\SIA e,#1,(\veroparagrafo.\veraformula)
 \global\advance\numfor by 1
\write15{\string\FU (#1){\equ(#1)}}
\9{ \write16{ EQ \equ(#1) == #1  }}}
\def \FU(#1)#2{\SIA fu,#1,#2 }
\def\etichettaa(#1){(A\veroparagrafo.\veraformula)
 \SIA e,#1,(A\veroparagrafo.\veraformula)
 \global\advance\numfor by 1
\write15{\string\FU (#1){\equ(#1)}}
\9{ \write16{ EQ \equ(#1) == #1  }}}
\def \FU(#1)#2{\SIA fu,#1,#2 }
\def\tetichetta(#1){\veroparagrafo.\veroteorema
\SIA e,#1,{\veroparagrafo.\veroteorema}
\global\advance\numteo by1
\write15{\string\FU (#1){\equ(#1)}}%
\9{\write16{ EQ \equ(#1) == #1}}}
\def\tetichettaa(#1){A\veroparagrafo.\veroteorema
\SIA e,#1,{A\veroparagrafo.\veroteorema}
\global\advance\numteo by1
\write15{\string\FU (#1){\equ(#1)}}%
\9{\write16{ EQ \equ(#1) == #1}}}
\def\letichetta(#1){\veroparagrafo.\verolemma
\SIA e,#1,{\veroparagrafo.\verolemma}
\global\advance\numlem by1
\write15{\string\FU (#1){\equ(#1)}}%
\9{\write16{ EQ \equ(#1) == #1}}}
\def\getichetta(#1){
 \SIA e,#1,{\verafigura}
 \global\advance\numfig by 1
\write15{\string\FU (#1){\equ(#1)}}
\9{ \write16{ Fig. \equ(#1) ha simbolo  #1  }}}

\def\veroparagrafo{\number\numsec}\def\veraformula{\number\numfor}
\def\verafigura{\number\numfig}\def\veroteorema{\number\numteo}
\def\verolemma{\number\numlem}

\def\geq(#1){\getichetta(#1)\galato(#1)}
\def\Eq(#1){\eqno{\etichetta(#1)\alato(#1)}}
\def\eq(#1){&\etichetta(#1)\alato(#1)}
\def\Eqa(#1){\eqno{\etichettaa(#1)\alato(#1)}}
\def\eqa(#1){&\etichettaa(#1)\alato(#1)}
\def\teq(#1){\tetichetta(#1)\talato(#1)}
\def\teqa(#1){\tetichettaa(#1)\talato(#1)}
\def\leq(#1){\letichetta(#1)\talato(#1)}

\def\Eqr{\eqno(\veroparagrafo.\veraformula)\advance\numfor by 1}
\def\eqr{&(\veroparagrafo.\veraformula)\advance\numfor by 1}
\def\Eqar{\eqno(A\veroparagrafo.\veraformula)\advance\numfor by 1}
\def\eqar{&(A\veroparagrafo.\veraformula)\advance\numfor by 1}

\def\eqv(#1){\senondefinito{fu#1}$\clubsuit$#1\write16{Manca #1 !}%
\else\csname fu#1\endcsname\fi}
\def\equ(#1){\senondefinito{e#1}\eqv(#1)\else\csname e#1\endcsname\fi}


\newdimen\gwidth

\def\commenta#1{\ifnum\bz=1\strut \vadjust{\kern-\profonditastruttura
 \vtop to \profonditastruttura{\baselineskip
 \profonditastruttura\vss
 \rlap{\kern\hsize\kern0.1truecm
  \vbox{\hsize=1.7truecm\raggedright\nota\noindent #1}}}}\fi}
\def\talato(#1){\ifnum\bz=1\strut \vadjust{\kern-\profonditastruttura
 \vtop to \profonditastruttura{\baselineskip
 \profonditastruttura\vss
 \rlap{\kern-1.2truecm{$\scriptstyle#1$}}}}\fi}
\def\alato(#1){\ifnum\bz=1
 {\vtop to \profonditastruttura{\baselineskip
 \profonditastruttura\vss
 \rlap{\kern-\hsize\kern-1.2truecm{$\scriptstyle#1$}}}}\fi}
\def\galato(#1){\ifnum\bz=1 \gwidth=\hsize 
 {\vtop to \profonditastruttura{\baselineskip
 \profonditastruttura\vss
 \rlap{\kern-\gwidth\kern-1.2truecm{$\scriptstyle#1$}}}}\fi}


\newskip\ttglue

\font\ftitolo=cmbx12 
\font\eighttt=cmtt8 \font\sevenit=cmti7  \font\sevensl=cmsl8
\font\sc=cmcsc10


\def\settepunti{\def\rm{\fam0\sevenrm}
\textfont0=\sevenrm \scriptfont0=\fiverm \scriptscriptfont0=\fiverm
\textfont1=\seveni \scriptfont1=\fivei   \scriptscriptfont1=\fivei
\textfont2=\sevensy \scriptfont2=\fivesy   \scriptscriptfont2=\fivesy
\textfont3=\tenex \scriptfont3=\tenex   \scriptscriptfont3=\tenex
\textfont\itfam=\sevenit  \def\it{\fam\itfam\sevenit}%
\textfont\slfam=\sevensl  \def\sl{\fam\slfam\sevensl}%
\textfont\ttfam=\eighttt  \def\tt{\fam\ttfam\eighttt}
\textfont\bffam=\sevenbf  \scriptfont\bffam=\fivebf
\scriptscriptfont\bffam=\fivebf  \def\bf{\fam\bffam\sevenbf}%
\tt \ttglue=.5em plus.25em minus.15em
\setbox\strutbox=\hbox{\vrule height6.5pt depth1.5pt width0pt}%
\normalbaselineskip=8pt\let\sc=\fiverm \normalbaselines\rm}

\let\nota=\settepunti


\font\tenmib=cmmib10
\font\sevenmib=cmmib10 scaled 800

\textfont5=\tenmib  \scriptfont5=\sevenmib  \scriptscriptfont5=\fivei

\mathchardef\aaa= "050B
\mathchardef\xxx= "0518
\mathchardef\oo = "0521
\mathchardef\Dp = "0540
\mathchardef\H  = "0548
\mathchardef\FFF= "0546
\mathchardef\ppp= "0570
\mathchardef\nnn= "0517

\newdimen\xshift \newdimen\xwidth \newdimen\yshift \newdimen\ywidth
\newdimen\laln

\def\ins#1#2#3{\nointerlineskip\vbox to0pt {\kern-#2 \hbox{\kern#1 #3}
\vss}}

\def\eqfig#1#2#3#4#5#6{
\xwidth=#1 \xshift=\hsize \advance\xshift 
by-\xwidth \divide\xshift by 2
\yshift=#2 \divide\yshift by 2
\midinsert
\parindent=0pt
\line{\hglue\xshift \vbox to #2{\vfil 
#3 \includegraphics{#4.ps}
}\hfill}
\nobreak
\*
\didascalia{\geq(#6)#5}\endinsert
}

\def\eqfigf#1#2#3#4#5#6{
\xwidth=#1 \xshift=\hsize \advance\xshift 
by-\xwidth \divide\xshift by 2
\yshift=#2 \divide\yshift by 2
\midinsert
\parindent=0pt
\line{\hglue\xshift \vbox to #2{\vfil 
#3 \includegraphics{#4.ps}
}\hfill}
\nobreak
\*
\didascalia{\geq(#6)#5}\endinsert
\setbox149\vbox{\unvbox149 \*\* \centerline{Fig. \equ(#6)} 
\nobreak
\*
\nobreak
\line{\hglue\xshift \vbox to #2{\vfil 
#3 \includegraphics{#4.ps}
}\hfill}\*}
\setbox150\vbox{\unvbox150 \parindent=0pt\* #5\*}
}

\def\eqfigbis#1#2#3#4#5#6#7{
\xwidth=#1 \multiply\xwidth by 2 
\xshift=\hsize \advance\xshift 
by-\xwidth \divide\xshift by 3
\yshift=#2 \divide\yshift by 2
\ywidth=#2
\line{\hfill
\vbox to \ywidth{\vfil #3 \includegraphics{#4.ps}}
\hglue20pt
\vbox to \ywidth{\vfil \includegraphics{#6.ps} #5}
\hfill\raise\yshift\hbox{#7}}}

\def\dimenfor#1#2{\par\xwidth=#1 \multiply\xwidth by 2 
\xshift=\hsize \advance\xshift 
by-\xwidth \divide\xshift by 3
\divide\xwidth by 2 
\yshift=#2 
\ywidth=#2}

\def\eqfigfor#1#2#3#4#5#6#7{
\midinsert
\parindent=0pt
\hbox to \hsize{\hskip\xshift 
\hbox to \xwidth{\vbox to \ywidth{\vfil#2\includegraphics{#10.ps}}\hfill}%
\hskip\xshift%
\hbox to \xwidth{\vbox to \ywidth{\vfil#3\includegraphics{#11.ps}}\hfill}\hfill}
\nobreak
\line{\hglue\xshift 
\hbox to \xwidth{\vbox to \ywidth{\vfil #4 \includegraphics{#12.ps}}\hfill}%
\hglue\xshift
\hbox to \xwidth{\vbox to\ywidth {\vfil #5 \includegraphics{#13.ps}}\hfill}\hfill}
\nobreak
\*\*
\didascalia{\geq(#7)#6}
\endinsert}

\def\eqfigforf#1#2#3#4#5#6#7{
\midinsert
\parindent=0pt
\hbox to \hsize{\hskip\xshift 
\hbox to \xwidth{\vbox to \ywidth{\vfil#2\includegraphics{#10.ps}}\hfill}%
\hskip\xshift%
\hbox to \xwidth{\vbox to \ywidth{\vfil#3\includegraphics{#11.ps}}\hfill}\hfill}
\nobreak
\line{\hglue\xshift 
\hbox to \xwidth{\vbox to \ywidth{\vfil #4 \includegraphics{#12.ps}}\hfill}%
\hglue\xshift
\hbox to \xwidth{\vbox to\ywidth {\vfil #5 \includegraphics{#13.ps}}\hfill}\hfill}
\nobreak
\*\*
\didascalia{\geq(#7)#6}
\endinsert
\setbox149\vbox{\unvbox149\* \centerline{Fig. \equ(#7)} \nobreak\* \nobreak
\*
\vbox{\hbox to \hsize{\hskip\xshift 
\hbox to \xwidth{\vbox to \ywidth{\vfil#2\includegraphics{#10.ps}}\hfill}%
\hskip\xshift%
\hbox to \xwidth{\vbox to \ywidth{\vfil#3\includegraphics{#11.ps}}\hfill}\hfill}
\nobreak
\line{\hglue\xshift 
\hbox to \xwidth{\vbox to \ywidth{\vfil #4 \includegraphics{#12.ps}}\hfill}%
\hglue\xshift
\hbox to \xwidth{\vbox to\ywidth {\vfil #5 \includegraphics{#13.ps}}\hfill}\hfill}
}\hfill}
\setbox150\vbox{\unvbox150\parindent=0pt\* #6\*}
}

\def\eqfigter#1#2#3#4#5#6#7{
\line{\hglue\xshift 
\vbox to \ywidth{\vfil #1 \includegraphics{#2.ps}}
\hglue30pt
\vbox to \ywidth{\vfil #3 \includegraphics{#4.ps}}\hfill}
\multiply\xshift by 3 \advance\xshift by \xwidth \divide\xshift by 2
\line{\hfill\hbox{#7}}
\line{\hglue\xshift 
\vbox to \ywidth{\vfil #5 \includegraphics{#6.ps}}}}


\def\7{\ifnum\modif=1\write13\else\write12\fi}
\def\8{\write13}


\def\gnuin #1 #2 #3 #4 #5 #6{\midinsert\vbox{\vbox to 260pt{
\hbox to 420pt{
\hbox to 200pt{\hfill\nota (a)\hfill}\hfill
\hbox to 200pt{\hfill\nota (b)\hfill}}
\vbox to 110pt{\vfill\hbox to 420pt{
\hbox to 200pt{\includegraphics{#1.ps}\hfill}\hfill
\hbox to 200pt{\includegraphics{#2.ps}\hfill}
}}\vfill
\hbox to 420pt{
\hbox to 200pt{\hfill\nota (c)\hfill}\hfill
\hbox to 200pt{\hfill\nota (d)\hfill}}
\vbox to 110pt{\vfill\hbox to 420pt{
\hbox to 200pt{\includegraphics{#3.ps}\hfill}\hfill
\hbox to 200pt{\includegraphics{#4.ps}\hfill}
}}\vfill}
\vskip0.25cm
\0\didascalia{\geq(#5): #6}}
\endinsert}

\def\gnuinf #1 #2 #3 #4 #5 #6{\midinsert\nointerlineskip\vbox to 260pt{
\hbox to 420pt{
\hbox to 200pt{\hfill\nota (a)\hfill}\hfill
\hbox to 200pt{\hfill\nota (b)\hfill}}
\vbox to 110pt{\vfill\hbox to 420pt{
\hbox to 200pt{\includegraphics{#1.ps}\hfill}\hfill
\hbox to 200pt{\includegraphics{#2.ps}\hfill}
}}\vfill
\hbox to 420pt{
\hbox to 200pt{\hfill\nota (c)\hfill}\hfill
\hbox to 200pt{\hfill\nota (d)\hfill}}
\vbox to 110pt{\vfill\hbox to 420pt{
\hbox to 200pt{\includegraphics{#3.ps}\hfill}\hfill
\hbox to 200pt{\includegraphics{#4.ps}\hfill}
}}\vfill}
\?
\0\didascalia{\geq(#5): #6}
\endinsert
\global\setbox150\vbox{\unvbox150 \*\*\0 Fig. \equ(#5): #6}
\global\setbox149\vbox{\unvbox149 \*\*
    \vbox{\centerline{Fig. \equ(#5)(a)} \nobreak
    \vbox to 200pt{\vfill\includegraphics{#1.ps_f}}}\*\*
    \vbox{\centerline{Fig. \equ(#5)(b)}\nobreak
    \vbox to 200pt{\vfill\includegraphics{#2.ps_f}}}\*\*
    \vbox{\centerline{Fig. \equ(#5)(c)}\nobreak
    \vbox to 200pt{\vfill\includegraphics{#3.ps_f}}}\*\*
    \vbox{\centerline{Fig. \equ(#5)(d)}\nobreak
    \vbox to 200pt{\vfill\includegraphics{#4.ps_f}}}
}}

\def\gnuins #1 #2 #3{\midinsert\nointerlineskip
\vbox{\line{\vbox to 220pt{\vfill
\includegraphics{#1.ps}}\hfill}
\*
\0\didascalia{\geq(#2): #3}}
\endinsert}

\def\gnuinsf #1 #2 #3{\midinsert\nointerlineskip
\vbox{\line{\vbox to 220pt{\vfill
\includegraphics{#1.ps}}\hfill}
\*
\0\didascalia{\geq(#2): #3}}\endinsert
\global\setbox150\vbox{\unvbox150 \*\*\0 Fig. \equ(#2): #3}
\global\setbox149\vbox{\unvbox149 \*\* 
    \vbox{\centerline{Fig. \equ(#2)}\nobreak
    \vbox to 220pt{\vfill\includegraphics{#1.ps_f}}}} 
}


\def\9#1{\ifnum\aux=1#1\else\relax\fi}
\let\numero=\number
\def\boh{\hbox{$\clubsuit$}\write16{Qualcosa di indefinito a pag. \the\pageno}}
\def\didascalia#1{\vbox{\nota\0#1\hfill}\vskip0.3truecm}
\def\frac#1#2{{#1\over #2}}
\def\V#1{{\underline{#1}}}
	\let\ciao=\bye
		\let\i=\infty
		
\def\media#1{\langle{#1}\rangle} 	\let\0=\noindent
\def\guida{\leaders\hbox to 1em{\hss.\hss}\hfill}
\def\tende#1{\vtop{\ialign{##\crcr\rightarrowfill\crcr
              \noalign{\kern-1pt\nointerlineskip}
              \hglue3.pt${\scriptstyle #1}$\hglue3.pt\crcr}}}
\def\otto{{\kern-1.truept\leftarrow\kern-5.truept\to\kern-1.truept}}

\def\={{ \; \equiv \; }}	\def\su{{\uparrow}}	\def\giu{{\downarrow}}
\ifnum\mgnf=0
    \def\openone{\leavevmode\hbox{\ninerm 1\kern-3.3pt\tenrm1}}%
\fi
\ifnum\mgnf=1
     \def\openone{\leavevmode\hbox{\ninerm 1\kern-3.6pt\tenrm1}}%
\fi

\def\2{{1\over2}}

\def\igb{
    \mathop{\raise4.pt\hbox{\vrule height0.2pt depth0.2pt width6.pt}
    \kern0.3pt\kern-9pt\int}}

\def\st{\scriptscriptstyle}
\let\\=\noindent
\def\*{\vskip0.5truecm}
\def\?{\vskip0.75truecm}
\def\item#1{\vskip0.1truecm\parindent=0pt\par\setbox0=\hbox{#1\ }
     \hangindent\wd0\hangafter 1 #1 \parindent=\stdindent}

\def\annota#1#2{\footnote{${}^#1$}{{\parindent0pt\\\settepunti#2\vfill}}}


\def\ie{\hbox{\sl i.e.\ }}
\def\eg{\hbox{\sl e.g.\ }}
\def\qed{\hfill\break\nobreak\vbox{\vglue.25truecm\line{\hfill\raise1pt 
          \hbox{\vrule height9pt width5pt depth0pt}}}\vglue.25truecm}


\def\gint(#1)(#2)(#3){{\cal D}#1^{#2}\,e^{(#1^{#2+},#3#1^{#2-})}}

  \def\V0{{\bf 0}}    \def\tt{{\bf t}}


\ifnum\cind=1
\def\prtindex#1{\immediate\write\indiceout{\string\parte{#1}{\the\pageno}}}
\def\capindex#1#2{\immediate\write\indiceout{\string\capitolo{#1}{#2}{\the\pageno}}}
\def\parindex#1#2{\immediate\write\indiceout{\string\paragrafo{#1}{#2}{\the\pageno}}}
\def\subindex#1#2{\immediate\write\indiceout{\string\sparagrafo{#1}{#2}{\the\pageno}}}
\def\appindex#1#2{\immediate\write\indiceout{\string\appendice{#1}{#2}{\the\pageno}}}
\def\paraindex#1#2{\immediate\write\indiceout{\string\paragrafoapp{#1}{#2}{\the\pageno}}}
\def\subaindex#1#2{\immediate\write\indiceout{\string\sparagrafoapp{#1}{#2}{\the\pageno}}}
\def\bibindex#1{\immediate\write\indiceout{\string\bibliografia{#1}{Bibliografia}{\the\pageno}}}
\def\preindex#1{\immediate\write\indiceout{\string\premessa{#1}{\the\pageno}}}
\else
\def\prtindex#1{}
\def\capindex#1#2{}
\def\parindex#1#2{}
\def\subindex#1#2{}
\def\appindex#1#2{}
\def\paraindex#1#2{}
\def\subaindex#1#2{}
\def\bibindex#1{}
\def\preindex#1{}
\fi

\def\leaderfill{\leaders\hbox to 1em{\hss . \hss} \hfill }


\newdimen\capsalto \capsalto=0pt
\newdimen\parsalto \parsalto=20pt
\newdimen\sparsalto \sparsalto=30pt
\newdimen\tratitoloepagina \tratitoloepagina=2\parsalto
\def\aboveparteskip{\bigskip \bigskip}
\def\belowparteskip{\medskip \medskip}
\def\abovecapitskip{\bigskip}
\def\belowcapitskip{\medskip}
\def\belowparskip{\smallskip}
%


\def\parte#1#2{
   \9{\immediate\write16
      {#1     pag.\numeropag{#2} }}
   \aboveparteskip 
   \noindent 
   {\ftitolo #1} 
   \hfill {\ftitolo \numeropag{#2}}\par
   \belowparteskip
   }


\def\premessa#1#2{
   \9{\immediate\write16
      {#1     pag.\numeropag{#2} }}
   \abovecapitskip 
   \noindent 
   {\it #1} 
   \hfill {\rm \numeropag{#2}}\par
   \belowcapitskip
   }


\def\bibliografia#1#2#3{
  \ifnum\stile=1
   \9{\immediate\write16
      {Bibliografia    pag.\numeropag{#3} }}
   \belowcapitskip
   \noindent 
   {\bf Bibliografia} 
   \hfill {\bf \numeropag{#3}}\par
  \else
    \paragrafo{#1}{References}{#3}
\fi
   }


\newdimen\newstrutboxheight
\newstrutboxheight=\baselineskip
\advance\newstrutboxheight by -\dp\strutbox
\newdimen\newstrutboxdepth
\newstrutboxdepth=\dp\strutbox
\newbox\newstrutbox
\setbox\newstrutbox = \hbox{\vrule 
   height \newstrutboxheight 
   width 0pt 
   depth \newstrutboxdepth 
   }
\def\newstrut
   {\relax \ifmmode \copy \newstrutbox \else \unhcopy \newstrutbox \fi}
%
%
\vfuzz=3.5pt
%
%
\newdimen\indexsize \indexsize=\hsize
\advance \indexsize by -\tratitoloepagina
\newdimen\dummy
\newbox\parnum
\newbox\parbody
\newbox\parpage
%

%

\def\mastercap#1#2#3#4#5{
   \9{\immediate\write16
      {Cap. #3:#4     pag.\numeropag{#5} }}
   \abovecapitskip
   \setbox\parnum=\hbox {\kern#1\newstrut{#2}
			{\bf Capitolo~\number#3.}~}
   \dummy=\indexsize
   \advance\indexsize by -\wd\parnum
   \setbox\parbody=\vbox {
      \hsize = \indexsize \noindent \newstrut 
      {\bf #4}
      \newstrut \hss}
   \indexsize=\dummy
   \setbox\parnum=\vbox to \ht\parbody {
      \box\parnum
      \vfill 
      }
   \setbox\parpage = \hbox to \tratitoloepagina {
      \hss {\bf \numeropag{#5}}}
   \noindent \box\parnum\box\parbody\box\parpage\par
   \belowcapitskip
   }
\def\capitolo#1#2#3{\mastercap{\capsalto}{}{#1}{#2}{#3}}
%


%
\def\masterapp#1#2#3#4#5{
   \9{\immediate\write16
      {App. #3:#4     pag.\numeropag{#5} }}
   \abovecapitskip
   \setbox\parnum=\hbox {\kern#1\newstrut{#2}
			{\bf Appendice~A\number#3:}~}
   \dummy=\indexsize
   \advance\indexsize by -\wd\parnum
   \setbox\parbody=\vbox {
      \hsize = \indexsize \noindent \newstrut 
      {\bf #4}
      \newstrut \hss}
   \indexsize=\dummy
   \setbox\parnum=\vbox to \ht\parbody {
      \box\parnum
      \vfill 
      }
   \setbox\parpage = \hbox to \tratitoloepagina {
      \hss {\bf \numeropag{#5}}}
   \noindent \box\parnum\box\parbody\box\parpage\par
   \belowcapitskip
   }
\def\appendice#1#2#3{\masterapp{\capsalto}{}{#1}{#2}{#3}}
%

%

\def\masterpar#1#2#3#4#5{
   \9{\immediate\write16
      {par. #3:#4     pag.\numeropag{#5} }}
   \setbox\parnum=\hbox {\kern#1\newstrut{#2}\number#3.~}
   \dummy=\indexsize
   \advance\indexsize by -\wd\parnum
   \setbox\parbody=\vbox {
      \hsize = \indexsize \noindent \newstrut 
      #4
      \newstrut \hss}
   \indexsize=\dummy
   \setbox\parnum=\vbox to \ht\parbody {
      \box\parnum
      \vfill 
      }
   \setbox\parpage = \hbox to \tratitoloepagina {
      \hss \numeropag{#5}}
   \noindent \box\parnum\box\parbody\box\parpage\par
   \belowparskip
   }
\def\paragrafo#1#2#3{\masterpar{\parsalto}{}{#1}{#2}{#3}}
\def\sparagrafo#1#2#3{\masterpar{\sparsalto}{}{#1}{#2}{#3}}
%


%
\def\masterpara#1#2#3#4#5{
   \9{\immediate\write16
      {par. #3:#4     pag.\numeropag{#5} }}
   \setbox\parnum=\hbox {\kern#1\newstrut{#2}A\number#3.~}
   \dummy=\indexsize
   \advance\indexsize by -\wd\parnum
   \setbox\parbody=\vbox {
      \hsize = \indexsize \noindent \newstrut 
      #4
      \newstrut \hss}
   \indexsize=\dummy
   \setbox\parnum=\vbox to \ht\parbody {
      \box\parnum
      \vfill 
      }
   \setbox\parpage = \hbox to \tratitoloepagina {
      \hss \numeropag{#5}}
   \noindent \box\parnum\box\parbody\box\parpage\par
   \belowparskip
   }
\def\paragrafoapp#1#2#3{\masterpara{\parsalto}{}{#1}{#2}{#3}}
\def\sparagrafoapp#1#2#3{\masterpara{\sparsalto}{}{#1}{#2}{#3}}
%


\ifnum\stile=1

\def\newcap#1{\setcap{#1}
\vskip2.truecm\advance\numsec by 1
\\{\ftitolo \numero\numsec. #1}
\capindex{\numero\numsec}{#1}
\vskip1.truecm\numfor=1\pgn=1\numpar=1\numteo=1\numlem=1
}

\def\newapp#1{\setcap{#1}
\vskip2.truecm\advance\numsec by 1
\\{\ftitolo A\numero\numsec. #1}
\appindex{A\numero\numsec}{#1}
\vskip1.truecm\numfor=1\pgn=1\numpar=1\numteo=1\numlem=1
}

\def\newpar#1{
\vskip1.truecm
\vbox{
\\{\bf \numero\numsec.\numero\numpar. #1}
\parindex{\numero\numsec.\numero\numpar}{#1}
\*{}}
\nobreak
\advance\numpar by 1
}

\def\newpara#1{
\vskip1.truecm
\vbox{
\\{\bf A\numero\numsec.\numero\numpar. #1}
\paraindex{\numero\numsec.\numero\numpar}{#1}
\*{}}
\nobreak
\advance\numpar by 1
}

\else

\def\newsec#1{\vskip1.truecm
\advance\numsec by 1
\vbox{
\\{\bf \numero\numsec. #1}
\parindex{\numero\numsec}{#1}
\*{}}\numfor=1\pgn=1\numpar=1\numteo=1\numlem=1
\nobreak
}

\def\newapp#1{\vskip1.truecm
\advance\numsec by 1
\vbox{
\\{\bf A\numero\numsec. #1}
\appindex{A\numero\numsec}{#1}
\*{}}\numfor=1\pgn=1\numpar=1\numteo=1\numlem=1
}

\def\biblio{\vskip1.truecm
\vbox{
\\{\bf References.}\*{}
\bibindex{{}}}\nobreak\makebiblio
}

\fi


\newread\indicein
\newwrite\indiceout

\def\faindice{
\openin\indicein=\jobname.ind
\ifeof\indicein\relax\else{
\ifnum\stile=1
  \pagenumbersind
  \pageno=-1
  \setind
  \null
  \vskip 2.truecm
  \\{\ftitolo Indice}
  \vskip 1.truecm
  \parskip = 0pt
  \input \jobname.ind
  \ppaginan
\else
\\{\bf Index}
\*{}
 \input \jobname.ind
\fi}\fi
\closein\indicein
\def\nomeindice{\jobname.ind}
\immediate\openout \indiceout = \nomeindice
}


\newwrite\bib
\immediate\openout\bib=\jobname.bib
\global\newcount\bibex
\bibex=0
\def\verabib{\number\bibex}

\ifnum\tipobib=0
\def\cita#1{\expandafter\ifx\csname c#1\endcsname\relax
\hbox{$\clubsuit$}#1\write16{Manca #1 !}%
\else\csname c#1\endcsname\fi}
\def\rife#1#2#3{\immediate\write\bib{\string\raf{#2}{#3}{#1}}
\immediate\write15{\string\C(#1){[#2]}}
\setbox199=\hbox{#2}\ifnum\maxit < \wd199 \maxit=\wd199\fi}
\else
\def\cita#1{%
\expandafter\ifx\csname d#1\endcsname\relax%
\expandafter\ifx\csname c#1\endcsname\relax%
\hbox{$\clubsuit$}#1\write16{Manca #1 !}%
\else\probib(ref. numero )(#1)%
\csname c#1\endcsname%
\fi\else\csname d#1\endcsname\fi}%
\def\rife#1#2#3{\immediate\write15{\string\Cp(#1){%
\string\immediate\string\write\string\bib{\string\string\string\raf%
{\string\verabib}{#3}{#1}}%
\string\Cn(#1){[\string\verabib]}%
\string\CCc(#1)%
}}}%
\fi

\def\Cn(#1)#2{\expandafter\xdef\csname d#1\endcsname{#2}}
\def\CCc(#1){\csname d#1\endcsname}
\def\probib(#1)(#2){\global\advance\bibex+1%
\9{\immediate\write16{#1\verabib => #2}}%
}

\def\C(#1)#2{\SIA c,#1,{#2}}
\def\Cp(#1)#2{\SIAnx c,#1,{#2}}

\def\SIAnx #1,#2,#3 {\senondefinito{#1#2}%
\expandafter\def\csname#1#2\endcsname{#3}\else%
\write16{???? ma #1,#2 e' gia' stato definito !!!!}\fi}

\bibskip=10truept
\def\hboxto{\hbox to}

\catcode`\{=12\catcode`\}=12
\catcode`\<=1\catcode`\>=2
\immediate\write\bib<
	\string\halign{\string\hboxto \string\maxit%
	{\string #\string\hfill}&%
        \string\vtop{\string\parindent=0pt\string\advance\string\hsize%
	by -1.55truecm%
	\string#\string\vskip \bibskip
	}\string\cr%
>
\catcode`\{=1\catcode`\}=2
\catcode`\<=12\catcode`\>=12

\def\raf#1#2#3{\ifnum \bz=0 [#1]&#2 \cr\else
\llap{${}_{\rm #3}$}[#1]&#2\cr\fi}

\newread\bibin

\catcode`\{=12\catcode`\}=12
\catcode`\<=1\catcode`\>=2
\def\chiudibib<
\catcode`\{=12\catcode`\}=12
\catcode`\<=1\catcode`\>=2
\immediate\write\bib<}>
\catcode`\{=1\catcode`\}=2
\catcode`\<=12\catcode`\>=12
>
\catcode`\{=1\catcode`\}=2
\catcode`\<=12\catcode`\>=12

\def\makebiblio{
\ifnum\tipobib=0
\advance \maxit by 10pt
\else
\maxit=1.truecm
\fi
\chiudibib
\immediate \closeout\bib
\openin\bibin=\jobname.bib
\ifeof\bibin\relax\else
\raggedbottom
\input \jobname.bib
\fi
}

\openin13=#1.aux \ifeof13 \relax \else
\input #1.aux \closein13\fi
\openin14=\jobname.aux \ifeof14 \relax \else
\input \jobname.aux \closein14 \fi
\immediate\openout15=\jobname.aux

\titolo{(Global and Local) Fluctuations of Phase Space Contraction in
Deterministic Stationary Non-equilibrium}

\autore{F. Bonetto}{Department of Mathematics, Rutgers University,
New Brunswick, NJ 08903}
\autore{N.I. Chernov}{Department of Mathematics, University of Alabama
in Birmingham, Birmingham, AL 35249}
\autore{J.L. Lebowitz}{Department of Mathematics and Physics,
Rutgers University, New Brunswick, NJ 08903}

\abstract{We studied numerically the validity of the fluctuation theorem
introduced in \cita{ECM2}\cita{GC} for a 2-dimensional system of particles
maintained in a steady shear flow by Maxwell daemon boundary conditions
\cita{CL}. The theorem was found to hold if one considers the total
phase space contraction $\sigma$ occuring at collisions with both
walls: $\sigma=\sigma^\su+\sigma^\giu$.  An attempt to extend it to
more local quantities $\sigma^\su$ and $\sigma^\giu$, corresponding to
the collisions with the top or bottom wall only, gave negative
results. The time decay of the correlations in $\sigma^{\su,\giu}$ was
very slow compared to that of $\sigma$.}

\parole{dynamical system, time reversibility, fluctuation theorem,
local phase space contraction}

\prima

\newsec{Introduction}

The microscopic structure of systems through which there is transport
of energy or momentum, is a central problem in non-equilibrium
statistical mechanics. For gases a partial answer to this question (on
the mesoscopic/kinetic level) is provided by the stationary
solution of the Boltzmann equation with suitable, \eg Maxwellian,
boundary conditions. Going beyond kinetic theory has proven to be very
difficult and we still lack a full understanding of stationary
non-equilibrium states (SNS) on the microscopic level.

This gap goes beyond that of computational complexity or even of
technical difficulty in proving the validity of the formulas derived
formally, something already present also in the kinetic theory and in
equilibrium statistical mechanics. What is still missing, at the
present time, are well defined formal procedure which would, at least
in principle, provide answers to questions of physical relevance for
SNS of macroscopic systems. Thus, we have no ``statistical mechanical
formula'' for the decay of spatial or time displaced correlation
functions in a SNS. In particular, we have no {\it a priori} formalism
for deciding on the slow power law decay of the spatial correlations
predicted by computer simulations, fluctuating hydrodynamics and
confirmed by experiments \cita{EE}. We also have no formula for
computing the stationary heat flux through a metal or plasma or the
momentum flux through a fluid which goes beyond linear response
theory.  Such a formula should, like the Green-Kubo formula for linear
transport, depend only on on the internal Hamiltonian of the system
and the impressed macroscopic constraints driving the system, \eg a
specified temperature or velocity field on the boundaries of the
system.
  
While the usual modeling of such systems has been via stochastic
boundaries \cita{ST} there has been much interest recently in the study
of SNS of particle systems evolving via entirely deterministic
dynamics in the hope that dynamical systems theory will provide new
insights into non-equilibrium behavior \cita{AAVV}\cita{BB}. As is well known
the existence of such SNS in finite systems is incompatible with the
usual Hamiltonian evolution, believed to accurately describe the
dynamics of systems in which quantum effects are unimportant. For such
dynamics the only realizable stationary states are those which depend
solely on the global constants of motion. In realistic systems these
are just the total energy and (under suitable boundary conditions) the
total momentum and angular momentum.

Adding external (globally non-gradient) forces to the dynamics, \eg a
uniform field in a system with periodic boundary conditions (such as
would arise from a changing magnetic flux crossing the surface bounded
by a conductor) will typically result in the system gaining energy
continuously from the field, excluding the possibility of SNS.  This
necessitates the use of non phase space volume conserving forces for
modeling SNS. Various models of such dynamics have been investigated
through computer simulations and heuristic analysis \cita{BB}, and
there have also been some mathematical results motivated by physical
considerations of such SNS \cita{GC}. There is however, as already
mentioned, still a very wide gap between the mathematical and physical
results obtained from these models and a full theory of macroscopic
SNS.

While it is not clear at present whether and how the dynamical system
approach will answer such questions it seems reasonable to explore the
behavior of such SNS. This is particularly so for models in which the
imposed, model dependent, driving and thermostating terms are confined
to the boundaries of the system while the dynamics in the interior of
the system remain realistically Hamiltonian. Several such models of
shear flow were investigated via computer simulations and some
heuristic analysis in \cita{CL}. Here we continue our investigation
focusing on the behavior of the fluctuation in the phase space volume
contraction $\sigma$ which occurs at the boundaries. $\sigma$ was found in
\cita{CL} to be approximately equal, when the size of the system is
sufficiently large for it to be in local thermal equilibrium, to the
hydrodynamic entropy production inside the system. 
The same quantity, in a different model of shear flow, was studied in
\cita{ECM2} where an interesting relation for its large deviations was
experimentally found.  This relation is now a rigorous result for
large deviations of phase space contraction under suitable conditions
on the dynamics \cita{GC}. Whether the results hold when the
conditions are not satisfied exactly is a question of great
relevance. The exploration of this and related quantities is the
subject of the present work.

In the next section we describe the model and the check of the
Gallavotti-Cohen result \cita{GC} for our system. In section 3 we
investigate fluctuations and time dependent correlations of the volume
contractions $\sigma^\su$ and $\sigma^\giu$ produced by collisions
with the top and bottom walls.
\vfill\eject
\newsec{Numerical simulations: the fluctuation theorem}

\0{\it A: The model}
\nobreak\*{}\nobreak
We consider the 2-dimensional shear flow model first introduced in
\cita{CL} and further studied in \cita{PD}: $N$ identical disks of
radius $r$ evolve in the interior of the system according to
Hamiltonian dynamics with hard core interaction among them while
Maxwell deamons at the walls drive the system away from
equilibrium. More specifically the particles move on the surface of a
cylinder (\ie the system has periodic boundary conditions on the
vertical sides) with reflection rules at the top and bottom walls
simulating macroscopic moving walls. The rules are as follows: when a
particle collides with the upper wall, making an angle $\phi$ between
the positive $x$-direction and the incoming velocity then the outgoing
velocity angle $\psi$ will be given by a (to be specified) reflection
rule $f$:

$$\psi=f(\phi)\Eq(ref)$$
  A similar rule applies on the lower wall with the only difference
that $\psi$ and $\phi$ refer to the angle between the incoming and
outgoing velocity and the negative $x$-direction. Since the modulus of
the velocity is preserved during a collision the total kinetic energy
of the system is a constant of the motion.

We will assume that $f$ is ``time reversible'', \ie it
satisfies $\phi=f(\pi-f(\pi-\phi))$. Observe that if this condition
holds the system is time reversible: if one inverts the velocities
of all the particle the system traces back its past trajectory. In the
numerical simulations we will consider one of the reflection rules
introduced in \cita{CL}, namely:

$$\psi=(\pi+b)-\sqrt{(\pi+b)^2-\phi(\phi+2b)}\Eq(refb)$$
where $b$ is used to modulate the intensity of the shear with $b=\i$
representing elastic collisions (see Fig. \equ(fig1)).

\eqfig{200pt}{100pt}{
\ins{70pt}{96pt}{$\phi$}
\ins{110pt}{96pt}{$\psi=f(\phi)$}
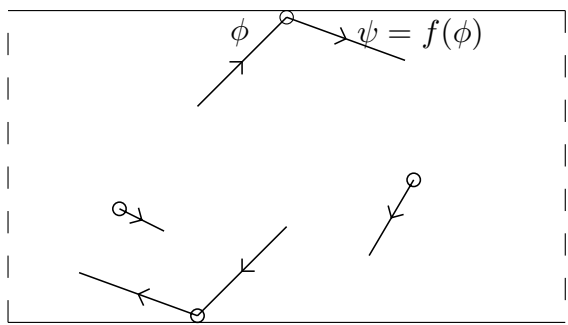
}{sys}{Fig. 1: Schematic representation of the dynamics of
the system}{fig1}

During a collision with the walls the Liouville measure is not
conserved. In fact at each ``reflection'' there is a phase space
contraction $\alpha$ equal to \cita{CL}: 

$$\alpha=-\log\left(\frac{\sin\psi}{\sin\phi}f'(\phi)\right)\Eq(con)$$

It is natural to consider the collisions with the walls as timing
events for the system, \ie we can consider the function $T$ that
associates to any point $X$ in the phase space of the system, \ie the
energy surface $\Sigma$, the first time at which a particle collides
with a wall and define the map, $S:\Sigma\to\tilde\Sigma$,
$S(X)=F(\Phi(X,T(X)))$, $\tilde \Sigma$ is the set of points at which
at least one particle is colliding with the walls where $\Phi(X,t)$ is
the point $X_t$ obtained from $X$ via the flow generated by the
dynamics and $F:\tilde \Sigma\to\tilde \Sigma$ is the collision map,
\ie the map that change the velocity of the colliding particle
according to the reflection rule $f$.

Given a point $X$ we define 

$$\alpha_\tau(X)=\sum_{i=-[\tau/2]}^{[\tau/2]-1}\alpha(S^i(X))\Eq(atau)$$
and
$$\sigma_\tau(X)=\frac{\alpha_\tau(X)}{\media{\alpha_\tau}_+}\Eq(stau)$$
where $[t]$ is the biggest integer less that $t$, $\media{\cdot}_+$
refer to the mean with respect to the forward invariant distribution.
Call $\pi_\tau(p)$ the distribution function of
$\sigma_\tau$, and let

$$x_\tau(p)=\frac{1}{\media{\alpha_\tau}_+}
\log\frac{\pi_\tau(p)}{\pi_\tau(-p)}\Eq(flu)$$
then the chaotic hypothesis of \cita{GC} implies that
$$\lim_{\tau\to\infty}x_\tau(p)=p\Eq(flu1)$$

To check numerically this relation numerical simulations were carried
out on systems of $N=10,20,30$ and $40$ particles of radius $R=1$ in a
``square'' box of size $L=\sqrt{dN}$, where $d=N/L^2$ is the number
density kept constant to $d=3.4\times 10^{-2}$, and with the parameter
$b=25$.  The initial configuration was chosen randomly in such a way
that the energy per particle
$e_0=\frac{1}{2N}\sum_{i=0}^{N}v_i^2=0.5$. In the next table we give
some of the interesting dynamical quantities associated with the
system.
\vbox{
$$\vcenter{
\halign{\strut\vrule\kern 3mm # \hfill\kern 3mm&\vrule \kern 3mm $#$ \kern
3mm&\vrule\kern 3mm $#$ \kern3mm&\vrule \kern 3mm $#$ \kern3mm\vrule&\kern
3mm $#$ \kern3mm\vrule\cr
\noalign{\hrule}
&10&20&30&40\cr
\noalign{\hrule}
$d$ &3.4\times 10^{-2}&3.4\times 10^{-2}&3.4\times 10^{-2}&\cr
\noalign{\hrule}
$t_0$&20.6&30.6&38.1&44.4\cr
\noalign{\hrule}
$\nu$ &2.11&3.19&4.00&4.67\cr
\noalign{\hrule}
$\langle \alpha\rangle_+$&1.24\times 10^{-2}&1.68\times 10^{-2}&2.00\times 10^{-2}&2.26\times 10^{-2}\cr
\noalign{\hrule}
$\l_{max}$&0.92&0.74&0.63&0.57\cr
\noalign{\hrule}}}$$
\nobreak
\didascalia{Table 1: static and dynamic quantity for the simulation
where p.s.c stands for phase space contraction.}
}

Here $t_0$ is the mean time between successive collisions with the
walls for a given particle, $\nu$ is the number of binary collisions
(\ie collisions between two particles) between two consecutive collisions
with the walls for a given particle and $\langle\alpha\rangle_+$ in
the mean phase space contraction rate. Finally we give the maximum
Lyapunov exponent of the map $S$. The data, given without error estimates,
are intended to give a rough image of the dynamics.
\nobreak
\*
\nobreak
\0{\it B: The fluctuation law}
\nobreak
\* 
\nobreak
The check of the fluctuation theorem was done as in \cita{BGG}. A long
trajectory of $5\times10^8$ collisions with the walls was simulated
and the phase space contraction was recorded for every 100
collisions. The main difference with \cita{BGG} is that we did not
attempt to decorrelate the adjacent data segment by leaving out a
fixed number of collision between them. In fact in \cita{BGG} this was
possible because the self-correlation of the phase space contraction
was decaying rapidly enough to leave out just few collisions. In the
present system we find that while the total phase space contraction
rate has rapidly decaying correlation the partial ones (to be defined
precisely and studied in the next section) have a very slow
decay. Hence we would have to discard too many collisions to
decorrelate the adjacent data segments and we therefore decided to
discard no collision also for the total phase space contraction to
have a consistent analysis.

In Fig. \equ(fig2) we show the graph of $x_\tau(p)$ for $N=20$ and
several value of $\tau$. 

\dimen2=25pt \advance\dimen2 by 158pt
\dimen3=196pt
\dimenfor{\dimen3}{199pt}
\eqfigfor{fig_tot}{
\ins{176pt}{0pt}{$\hbox to20pt{\hfill ${\scriptstyle p}$\hfill}$}
\ins{12pt}{\dimen2}{$\hbox to25pt{\hfill ${\scriptstyle x(p)}$\hfill}$}
\ins{99pt}{176pt}{$\hbox to79pt{\hfill ${\scriptstyle \tau=100}$\hfill}$}
\ins{8pt}{3pt}{$\hbox to 30pt{\hfill${\scriptstyle 0}$\hfill}$}
\ins{38pt}{3pt}{$\hbox to 30pt{\hfill${\scriptstyle 1}$\hfill}$}
\ins{68pt}{3pt}{$\hbox to 30pt{\hfill${\scriptstyle 2}$\hfill}$}
\ins{98pt}{3pt}{$\hbox to 30pt{\hfill${\scriptstyle 3}$\hfill}$}
\ins{128pt}{3pt}{$\hbox to 30pt{\hfill${\scriptstyle 4}$\hfill}$}
\ins{158pt}{3pt}{$\hbox to 30pt{\hfill${\scriptstyle 5}$\hfill}$}
\ins{0pt}{20pt}{$\hbox to 14pt{\hfill${\scriptstyle 0}$}$}
\ins{0pt}{41pt}{$\hbox to 14pt{\hfill${\scriptstyle 1}$}$}
\ins{0pt}{62pt}{$\hbox to 14pt{\hfill${\scriptstyle 2}$}$}
\ins{0pt}{83pt}{$\hbox to 14pt{\hfill${\scriptstyle 3}$}$}
\ins{0pt}{105pt}{$\hbox to 14pt{\hfill${\scriptstyle 4}$}$}
\ins{0pt}{126pt}{$\hbox to 14pt{\hfill${\scriptstyle 5}$}$}
\ins{0pt}{147pt}{$\hbox to 14pt{\hfill${\scriptstyle 6}$}$}
\ins{0pt}{168pt}{$\hbox to 14pt{\hfill${\scriptstyle 7}$}$}
}{\ins{176pt}{0pt}{$\hbox to20pt{\hfill ${\scriptstyle p}$\hfill}$}
\ins{12pt}{\dimen2}{$\hbox to25pt{\hfill ${\scriptstyle x(p)}$\hfill}$}
\ins{99pt}{176pt}{$\hbox to79pt{\hfill ${\scriptstyle \tau=400}$\hfill}$}
\ins{-9pt}{3pt}{$\hbox to 55pt{\hfill${\scriptstyle 0.0}$\hfill}$}
\ins{47pt}{3pt}{$\hbox to 55pt{\hfill${\scriptstyle 0.5}$\hfill}$}
\ins{102pt}{3pt}{$\hbox to 55pt{\hfill${\scriptstyle 1.0}$\hfill}$}
\ins{0pt}{16pt}{$\hbox to 14pt{\hfill${\scriptstyle 0.0}$}$}
\ins{0pt}{64pt}{$\hbox to 14pt{\hfill${\scriptstyle 0.5}$}$}
\ins{0pt}{112pt}{$\hbox to 14pt{\hfill${\scriptstyle 1.0}$}$}
\ins{0pt}{160pt}{$\hbox to 14pt{\hfill${\scriptstyle 1.5}$}$}
}{\ins{176pt}{0pt}{$\hbox to20pt{\hfill ${\scriptstyle p}$\hfill}$}
\ins{12pt}{\dimen2}{$\hbox to25pt{\hfill ${\scriptstyle x(p)}$\hfill}$}
\ins{99pt}{176pt}{$\hbox to79pt{\hfill ${\scriptstyle \tau=600}$\hfill}$}
\ins{-29pt}{3pt}{$\hbox to 90pt{\hfill${\scriptstyle 0.0}$\hfill}$}
\ins{61pt}{3pt}{$\hbox to 90pt{\hfill${\scriptstyle 0.5}$\hfill}$}
\ins{0pt}{13pt}{$\hbox to 14pt{\hfill${\scriptstyle 0.0}$}$}
\ins{0pt}{95pt}{$\hbox to 14pt{\hfill${\scriptstyle 0.5}$}$}
}{\ins{176pt}{0pt}{$\hbox to20pt{\hfill ${\scriptstyle p}$\hfill}$}
\ins{12pt}{\dimen2}{$\hbox to25pt{\hfill ${\scriptstyle x(p)}$\hfill}$}
\ins{99pt}{176pt}{$\hbox to79pt{\hfill ${\scriptstyle \tau=800}$\hfill}$}
\ins{-11pt}{3pt}{$\hbox to 48pt{\hfill${\scriptstyle 0.0}$\hfill}$}
\ins{37pt}{3pt}{$\hbox to 48pt{\hfill${\scriptstyle 0.2}$\hfill}$}
\ins{85pt}{3pt}{$\hbox to 48pt{\hfill${\scriptstyle 0.4}$\hfill}$}
\ins{133pt}{3pt}{$\hbox to 48pt{\hfill${\scriptstyle 0.6}$\hfill}$}
\ins{0pt}{13pt}{$\hbox to 14pt{\hfill${\scriptstyle 0.0}$}$}
\ins{0pt}{50pt}{$\hbox to 14pt{\hfill${\scriptstyle 0.2}$}$}
\ins{0pt}{87pt}{$\hbox to 14pt{\hfill${\scriptstyle 0.4}$}$}
\ins{0pt}{124pt}{$\hbox to 14pt{\hfill${\scriptstyle 0.6}$}$}
\ins{0pt}{162pt}{$\hbox to 14pt{\hfill${\scriptstyle 0.8}$}$}
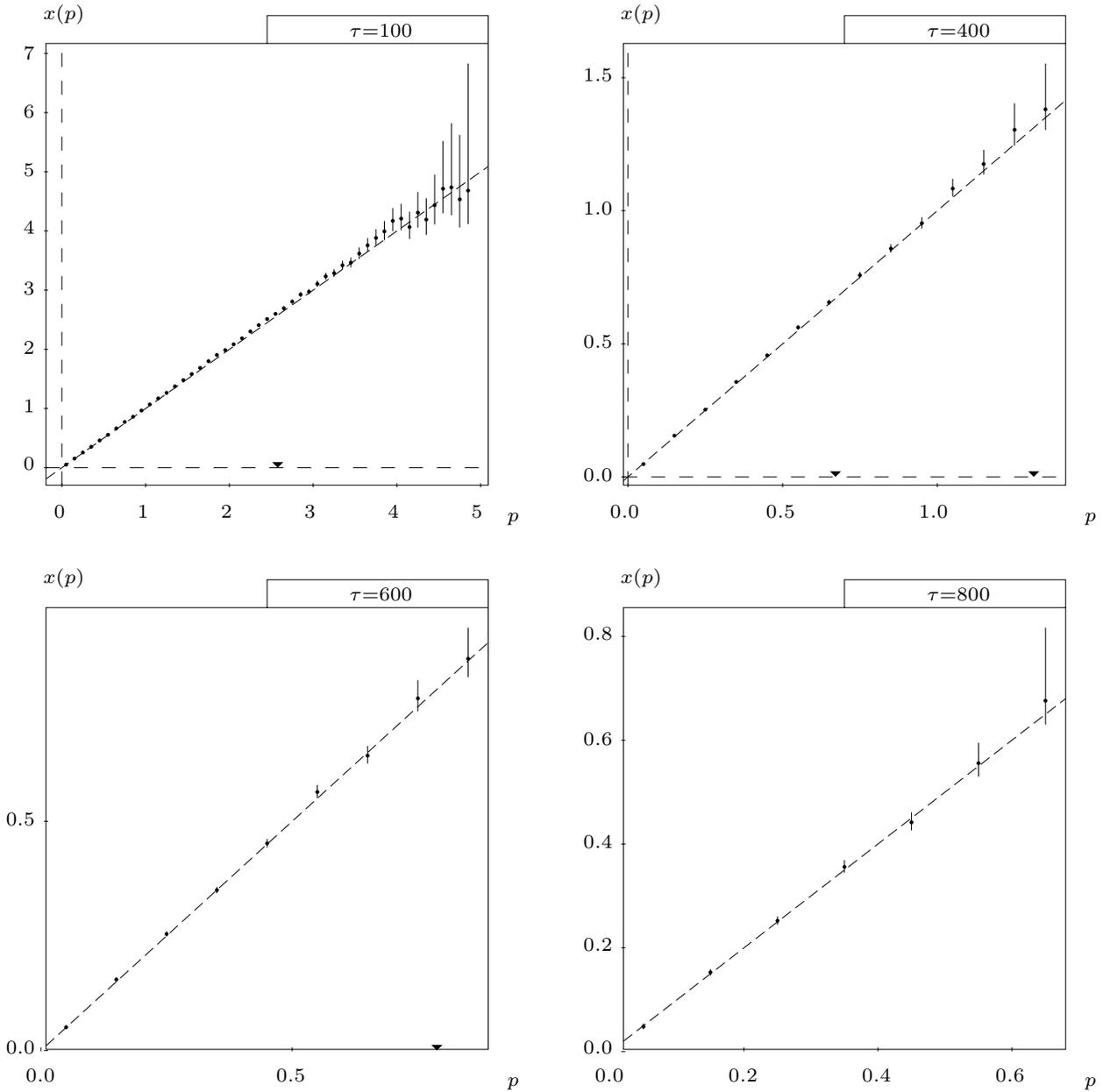
}{Fig. \equ(fig2): Graphs of $x_\tau(p)$ for
$\tau=100,400,600,800$ for $N=20$. The arrows represents the value 1 
plus or minus the standard deviation of $\sigma_\tau$.}{fig2}

As we can see already for $\tau=400$ the agreement between the
theoretical prediction and the experiment is very good.

To better follow the behavior of the fluctuation we have constructed
the function $x_\tau(p)$ for $N=10,20,30$ and $40$ for several value
of $\tau$ from 1 to 1000. We have then used a least square fit to fit
the experimental data with a law of the form $x_\tau(p)=\chi_\tau
p$. We chose a one parameter fit because $x_\tau(0)\equiv 0$. The
result are shown in Fig. \equ(fig3). As can be easily seen the evaluated
$\chi_\tau$ contain 1 within their error-bars starting from quite
small value of $\tau$. Moreover, one appears to be the asympotical
value of $\chi_\tau$.

\dimen2=25pt \advance\dimen2 by 126pt
\dimen3=204pt
\dimenfor{\dimen3}{166pt}
\eqfigfor{fig_f}{
\ins{184pt}{0pt}{$\hbox to20pt{\hfill ${\scriptstyle \tau}$\hfill}$}
\ins{12pt}{\dimen2}{$\hbox to25pt{\hfill ${\scriptstyle \chi_\tau}$\hfill}$}
\ins{103pt}{144pt}{$\hbox to83pt{\hfill ${\scriptstyle N=10}$\hfill}$}
\ins{10pt}{3pt}{$\hbox to 30pt{\hfill${\scriptstyle 0}$\hfill}$}
\ins{40pt}{3pt}{$\hbox to 30pt{\hfill${\scriptstyle 200}$\hfill}$}
\ins{70pt}{3pt}{$\hbox to 30pt{\hfill${\scriptstyle 400}$\hfill}$}
\ins{101pt}{3pt}{$\hbox to 30pt{\hfill${\scriptstyle 600}$\hfill}$}
\ins{131pt}{3pt}{$\hbox to 30pt{\hfill${\scriptstyle 800}$\hfill}$}
\ins{161pt}{3pt}{$\hbox to 30pt{\hfill${\scriptstyle 1000}$\hfill}$}
\ins{0pt}{19pt}{$\hbox to 14pt{\hfill${\scriptstyle 0.9}$}$}
\ins{0pt}{43pt}{$\hbox to 14pt{\hfill${\scriptstyle 1.2}$}$}
\ins{0pt}{66pt}{$\hbox to 14pt{\hfill${\scriptstyle 1.5}$}$}
\ins{0pt}{90pt}{$\hbox to 14pt{\hfill${\scriptstyle 1.8}$}$}
\ins{0pt}{114pt}{$\hbox to 14pt{\hfill${\scriptstyle 2.1}$}$}
\ins{0pt}{137pt}{$\hbox to 14pt{\hfill${\scriptstyle 2.4}$}$}
}{\ins{184pt}{0pt}{$\hbox to20pt{\hfill ${\scriptstyle \tau}$\hfill}$}
\ins{12pt}{\dimen2}{$\hbox to25pt{\hfill ${\scriptstyle \chi_\tau}$\hfill}$}
\ins{103pt}{144pt}{$\hbox to83pt{\hfill ${\scriptstyle N=20}$\hfill}$}
\ins{10pt}{3pt}{$\hbox to 30pt{\hfill${\scriptstyle 0}$\hfill}$}
\ins{40pt}{3pt}{$\hbox to 30pt{\hfill${\scriptstyle 200}$\hfill}$}
\ins{70pt}{3pt}{$\hbox to 30pt{\hfill${\scriptstyle 400}$\hfill}$}
\ins{101pt}{3pt}{$\hbox to 30pt{\hfill${\scriptstyle 600}$\hfill}$}
\ins{131pt}{3pt}{$\hbox to 30pt{\hfill${\scriptstyle 800}$\hfill}$}
\ins{161pt}{3pt}{$\hbox to 30pt{\hfill${\scriptstyle 1000}$\hfill}$}
\ins{0pt}{20pt}{$\hbox to 14pt{\hfill${\scriptstyle 0.9}$}$}
\ins{0pt}{36pt}{$\hbox to 14pt{\hfill${\scriptstyle 1.2}$}$}
\ins{0pt}{52pt}{$\hbox to 14pt{\hfill${\scriptstyle 1.5}$}$}
\ins{0pt}{68pt}{$\hbox to 14pt{\hfill${\scriptstyle 1.8}$}$}
\ins{0pt}{84pt}{$\hbox to 14pt{\hfill${\scriptstyle 2.1}$}$}
\ins{0pt}{101pt}{$\hbox to 14pt{\hfill${\scriptstyle 2.4}$}$}
\ins{0pt}{117pt}{$\hbox to 14pt{\hfill${\scriptstyle 2.7}$}$}
\ins{0pt}{133pt}{$\hbox to 14pt{\hfill${\scriptstyle 3.0}$}$}
}{\ins{184pt}{0pt}{$\hbox to20pt{\hfill ${\scriptstyle \tau}$\hfill}$}
\ins{12pt}{\dimen2}{$\hbox to25pt{\hfill ${\scriptstyle \chi_\tau}$\hfill}$}
\ins{103pt}{144pt}{$\hbox to83pt{\hfill ${\scriptstyle N=30}$\hfill}$}
\ins{10pt}{3pt}{$\hbox to 30pt{\hfill${\scriptstyle 0}$\hfill}$}
\ins{40pt}{3pt}{$\hbox to 30pt{\hfill${\scriptstyle 200}$\hfill}$}
\ins{70pt}{3pt}{$\hbox to 30pt{\hfill${\scriptstyle 400}$\hfill}$}
\ins{101pt}{3pt}{$\hbox to 30pt{\hfill${\scriptstyle 600}$\hfill}$}
\ins{131pt}{3pt}{$\hbox to 30pt{\hfill${\scriptstyle 800}$\hfill}$}
\ins{161pt}{3pt}{$\hbox to 30pt{\hfill${\scriptstyle 1000}$\hfill}$}
\ins{0pt}{21pt}{$\hbox to 14pt{\hfill${\scriptstyle 0.9}$}$}
\ins{0pt}{34pt}{$\hbox to 14pt{\hfill${\scriptstyle 1.2}$}$}
\ins{0pt}{47pt}{$\hbox to 14pt{\hfill${\scriptstyle 1.5}$}$}
\ins{0pt}{60pt}{$\hbox to 14pt{\hfill${\scriptstyle 1.8}$}$}
\ins{0pt}{73pt}{$\hbox to 14pt{\hfill${\scriptstyle 2.1}$}$}
\ins{0pt}{86pt}{$\hbox to 14pt{\hfill${\scriptstyle 2.4}$}$}
\ins{0pt}{99pt}{$\hbox to 14pt{\hfill${\scriptstyle 2.7}$}$}
\ins{0pt}{112pt}{$\hbox to 14pt{\hfill${\scriptstyle 3.0}$}$}
\ins{0pt}{125pt}{$\hbox to 14pt{\hfill${\scriptstyle 3.3}$}$}
\ins{0pt}{138pt}{$\hbox to 14pt{\hfill${\scriptstyle 3.6}$}$}
}{\ins{184pt}{0pt}{$\hbox to20pt{\hfill ${\scriptstyle \tau}$\hfill}$}
\ins{12pt}{\dimen2}{$\hbox to25pt{\hfill ${\scriptstyle \chi_\tau}$\hfill}$}
\ins{103pt}{144pt}{$\hbox to83pt{\hfill ${\scriptstyle N=40}$\hfill}$}
\ins{8pt}{3pt}{$\hbox to 33pt{\hfill${\scriptstyle 0}$\hfill}$}
\ins{42pt}{3pt}{$\hbox to 33pt{\hfill${\scriptstyle 200}$\hfill}$}
\ins{75pt}{3pt}{$\hbox to 33pt{\hfill${\scriptstyle 400}$\hfill}$}
\ins{109pt}{3pt}{$\hbox to 33pt{\hfill${\scriptstyle 600}$\hfill}$}
\ins{143pt}{3pt}{$\hbox to 33pt{\hfill${\scriptstyle 800}$\hfill}$}
\ins{0pt}{22pt}{$\hbox to 14pt{\hfill${\scriptstyle 0.9}$}$}
\ins{0pt}{33pt}{$\hbox to 14pt{\hfill${\scriptstyle 1.2}$}$}
\ins{0pt}{44pt}{$\hbox to 14pt{\hfill${\scriptstyle 1.5}$}$}
\ins{0pt}{56pt}{$\hbox to 14pt{\hfill${\scriptstyle 1.8}$}$}
\ins{0pt}{67pt}{$\hbox to 14pt{\hfill${\scriptstyle 2.1}$}$}
\ins{0pt}{78pt}{$\hbox to 14pt{\hfill${\scriptstyle 2.4}$}$}
\ins{0pt}{89pt}{$\hbox to 14pt{\hfill${\scriptstyle 2.7}$}$}
\ins{0pt}{100pt}{$\hbox to 14pt{\hfill${\scriptstyle 3.0}$}$}
\ins{0pt}{112pt}{$\hbox to 14pt{\hfill${\scriptstyle 3.3}$}$}
\ins{0pt}{123pt}{$\hbox to 14pt{\hfill${\scriptstyle 3.6}$}$}
\ins{0pt}{134pt}{$\hbox to 14pt{\hfill${\scriptstyle 3.9}$}$}
}{Fig. \equ(fig3): Behavior of $\chi_\tau$ as a function of
$\tau$ for $N=10,20,30$ and 40.}{fig3}

The analysis of the fluctuation law requires the construction of the
whole distribution function $\pi_\tau(p)$. But as $\tau$ grows
big fluctuations become more and more improbable and it is impossible
to construct $x_\tau(p)$ for $\tau>1000$. To go further with $\tau$ we
can observe that, starting with $\tau\simeq 5$, the distribution
$\pi_\tau(p)$ look very much like a Gaussian. Observe that if we
assume that the system is Anosov the central limit theorem implies a
Gaussian behavior near the maximum of the distribution but the
observed agreement goes, in our opinion, beyond the prediction of the
central limit theorem. Nevertheless it must be noted that the
distribution can't be Gaussian because it is easy to observe that
$|\alpha_\tau(X)|$ is bounded. Fig \equ(fig4) show the comparison with
a Gaussian for $N=20$ and several values of $\tau$.

\dimen2=25pt \advance\dimen2 by 158pt
\dimen3=202pt
\dimenfor{\dimen3}{199pt}
\eqfigfor{fig_g}{
\ins{182pt}{0pt}{$\hbox to20pt{\hfill ${\scriptstyle p}$\hfill}$}
\ins{18pt}{\dimen2}{$\hbox to25pt{\hfill ${\scriptstyle \pi_\tau}$\hfill}$}
\ins{99pt}{176pt}{$\hbox to79pt{\hfill ${\scriptstyle \tau=5}$\hfill}$}
\ins{-21pt}{3pt}{$\hbox to 35pt{\hfill${\scriptstyle -18}$\hfill}$}
\ins{15pt}{3pt}{$\hbox to 35pt{\hfill${\scriptstyle -12}$\hfill}$}
\ins{50pt}{3pt}{$\hbox to 35pt{\hfill${\scriptstyle -6}$\hfill}$}
\ins{85pt}{3pt}{$\hbox to 35pt{\hfill${\scriptstyle 0}$\hfill}$}
\ins{120pt}{3pt}{$\hbox to 35pt{\hfill${\scriptstyle 6}$\hfill}$}
\ins{156pt}{3pt}{$\hbox to 35pt{\hfill${\scriptstyle 12}$\hfill}$}
\ins{0pt}{21pt}{$\hbox to 20pt{\hfill${\scriptstyle 0.00}$}$}
\ins{0pt}{55pt}{$\hbox to 20pt{\hfill${\scriptstyle 0.03}$}$}
\ins{0pt}{90pt}{$\hbox to 20pt{\hfill${\scriptstyle 0.06}$}$}
\ins{0pt}{125pt}{$\hbox to 20pt{\hfill${\scriptstyle 0.09}$}$}
\ins{0pt}{159pt}{$\hbox to 20pt{\hfill${\scriptstyle 0.12}$}$}
}{\ins{182pt}{0pt}{$\hbox to20pt{\hfill ${\scriptstyle p}$\hfill}$}
\ins{18pt}{\dimen2}{$\hbox to25pt{\hfill ${\scriptstyle \pi_\tau}$\hfill}$}
\ins{99pt}{176pt}{$\hbox to79pt{\hfill ${\scriptstyle \tau=50}$\hfill}$}
\ins{9pt}{3pt}{$\hbox to 33pt{\hfill${\scriptstyle -6}$\hfill}$}
\ins{42pt}{3pt}{$\hbox to 33pt{\hfill${\scriptstyle -3}$\hfill}$}
\ins{76pt}{3pt}{$\hbox to 33pt{\hfill${\scriptstyle 0}$\hfill}$}
\ins{109pt}{3pt}{$\hbox to 33pt{\hfill${\scriptstyle 3}$\hfill}$}
\ins{143pt}{3pt}{$\hbox to 33pt{\hfill${\scriptstyle 6}$\hfill}$}
\ins{0pt}{21pt}{$\hbox to 20pt{\hfill${\scriptstyle 0.0}$}$}
\ins{0pt}{71pt}{$\hbox to 20pt{\hfill${\scriptstyle 0.1}$}$}
\ins{0pt}{121pt}{$\hbox to 20pt{\hfill${\scriptstyle 0.2}$}$}
\ins{0pt}{171pt}{$\hbox to 20pt{\hfill${\scriptstyle 0.3}$}$}
}{\ins{182pt}{0pt}{$\hbox to20pt{\hfill ${\scriptstyle p}$\hfill}$}
\ins{18pt}{\dimen2}{$\hbox to25pt{\hfill ${\scriptstyle \pi_\tau}$\hfill}$}
\ins{99pt}{176pt}{$\hbox to79pt{\hfill ${\scriptstyle \tau=500}$\hfill}$}
\ins{-10pt}{3pt}{$\hbox to 32pt{\hfill${\scriptstyle -2}$\hfill}$}
\ins{23pt}{3pt}{$\hbox to 32pt{\hfill${\scriptstyle -1}$\hfill}$}
\ins{55pt}{3pt}{$\hbox to 32pt{\hfill${\scriptstyle 0}$\hfill}$}
\ins{88pt}{3pt}{$\hbox to 32pt{\hfill${\scriptstyle 1}$\hfill}$}
\ins{121pt}{3pt}{$\hbox to 32pt{\hfill${\scriptstyle 2}$\hfill}$}
\ins{153pt}{3pt}{$\hbox to 32pt{\hfill${\scriptstyle 3}$\hfill}$}
\ins{0pt}{21pt}{$\hbox to 20pt{\hfill${\scriptstyle 0.0}$}$}
\ins{0pt}{55pt}{$\hbox to 20pt{\hfill${\scriptstyle 0.2}$}$}
\ins{0pt}{90pt}{$\hbox to 20pt{\hfill${\scriptstyle 0.4}$}$}
\ins{0pt}{125pt}{$\hbox to 20pt{\hfill${\scriptstyle 0.6}$}$}
\ins{0pt}{160pt}{$\hbox to 20pt{\hfill${\scriptstyle 0.8}$}$}
}{\ins{182pt}{0pt}{$\hbox to20pt{\hfill ${\scriptstyle p}$\hfill}$}
\ins{18pt}{\dimen2}{$\hbox to25pt{\hfill ${\scriptstyle \pi_\tau}$\hfill}$}
\ins{99pt}{176pt}{$\hbox to79pt{\hfill ${\scriptstyle \tau=1000}$\hfill}$}
\ins{6pt}{3pt}{$\hbox to 36pt{\hfill${\scriptstyle -0.8}$\hfill}$}
\ins{42pt}{3pt}{$\hbox to 36pt{\hfill${\scriptstyle 0.0}$\hfill}$}
\ins{78pt}{3pt}{$\hbox to 36pt{\hfill${\scriptstyle 0.8}$\hfill}$}
\ins{114pt}{3pt}{$\hbox to 36pt{\hfill${\scriptstyle 1.6}$\hfill}$}
\ins{150pt}{3pt}{$\hbox to 36pt{\hfill${\scriptstyle 2.4}$\hfill}$}
\ins{0pt}{21pt}{$\hbox to 20pt{\hfill${\scriptstyle 0.0}$}$}
\ins{0pt}{46pt}{$\hbox to 20pt{\hfill${\scriptstyle 0.2}$}$}
\ins{0pt}{70pt}{$\hbox to 20pt{\hfill${\scriptstyle 0.4}$}$}
\ins{0pt}{95pt}{$\hbox to 20pt{\hfill${\scriptstyle 0.6}$}$}
\ins{0pt}{120pt}{$\hbox to 20pt{\hfill${\scriptstyle 0.8}$}$}
\ins{0pt}{145pt}{$\hbox to 20pt{\hfill${\scriptstyle 1.0}$}$}
\ins{0pt}{169pt}{$\hbox to 20pt{\hfill${\scriptstyle 1.2}$}$}
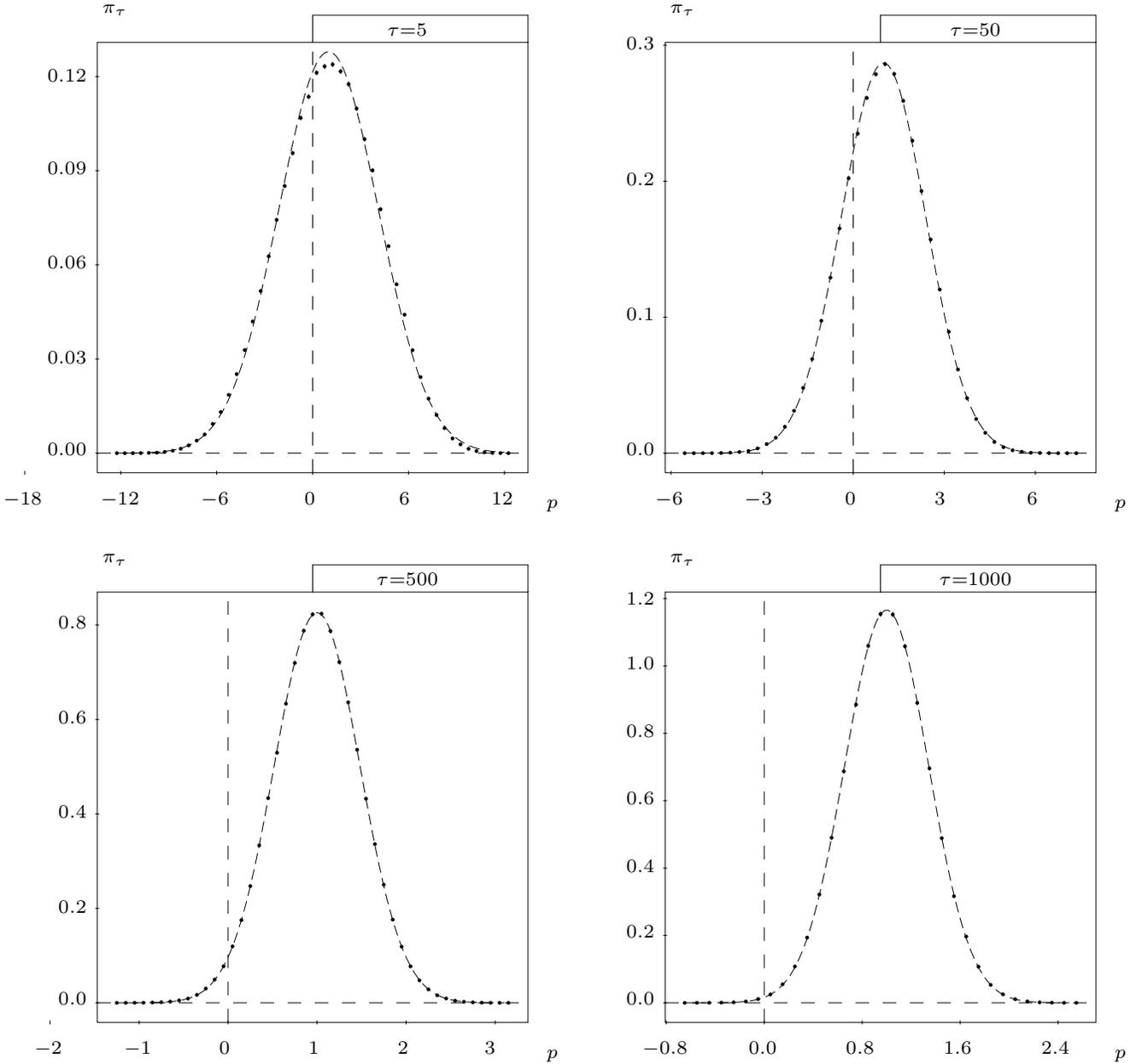
}{Fig. \equ(fig4): Comparison between the distribution $\pi_\tau(p)$,
for $N=20$ and several values of $\tau$, and a Gaussian with the same
variance and mean. The error-bars are smaller than the dimension of
the points.}{fig4}

Assuming that $\pi_\tau$ is a Gaussian immediately implies that
$x_\tau(p)$ is linear in $p$ and, calling $C_\tau$ the covariance of
$\sigma_\tau$, \ie $C_\tau=\langle
\sigma_\tau^2\rangle-1$,we have that 

$$\chi_\tau=\frac{2}{\langle\alpha_\tau\rangle C_\tau}\Eq(chi)$$

The covariance $C_\tau$ can be easily computed from the data and
permit us to go beyond the limit of $\tau=100$ met before. As a further
check we show the value of $\chi_\tau$ as computed from the best fit
and from the Gaussian hypothesis for small value of $\tau$ in
Fig. \equ(fig5).

\dimen2=25pt \advance\dimen2 by 198pt
\dimen3=354pt
\eqfig{\dimen3}{240pt}{
\ins{334pt}{0pt}{$\hbox to20pt{\hfill ${\scriptstyle \tau}$\hfill}$}
\ins{12pt}{\dimen2}{$\hbox to25pt{\hfill ${\scriptstyle \chi_\tau}$\hfill}$}
\ins{178pt}{216pt}{$\hbox to158pt{\hfill ${\scriptstyle N=30}$\hfill}$}
\ins{-15pt}{3pt}{$\hbox to 87pt{\hfill${\scriptstyle 0}$\hfill}$}
\ins{73pt}{3pt}{$\hbox to 87pt{\hfill${\scriptstyle 30}$\hfill}$}
\ins{160pt}{3pt}{$\hbox to 87pt{\hfill${\scriptstyle 60}$\hfill}$}
\ins{247pt}{3pt}{$\hbox to 87pt{\hfill${\scriptstyle 90}$\hfill}$}
\ins{0pt}{23pt}{$\hbox to 14pt{\hfill${\scriptstyle 1.0}$}$}
\ins{0pt}{45pt}{$\hbox to 14pt{\hfill${\scriptstyle 1.3}$}$}
\ins{0pt}{68pt}{$\hbox to 14pt{\hfill${\scriptstyle 1.6}$}$}
\ins{0pt}{90pt}{$\hbox to 14pt{\hfill${\scriptstyle 1.9}$}$}
\ins{0pt}{113pt}{$\hbox to 14pt{\hfill${\scriptstyle 2.2}$}$}
\ins{0pt}{135pt}{$\hbox to 14pt{\hfill${\scriptstyle 2.5}$}$}
\ins{0pt}{158pt}{$\hbox to 14pt{\hfill${\scriptstyle 2.8}$}$}
\ins{0pt}{180pt}{$\hbox to 14pt{\hfill${\scriptstyle 3.1}$}$}
\ins{0pt}{202pt}{$\hbox to 14pt{\hfill${\scriptstyle 3.4}$}$}
\laln=316pt
\setbox180=\hbox{\settepunti Fit}
\advance \laln by -\wd180
\ins{\laln}{188pt}{\copy180}
\laln=316pt
\setbox180=\hbox{\settepunti Gaussian}
\advance \laln by -\wd180
\ins{\laln}{176pt}{\copy180}
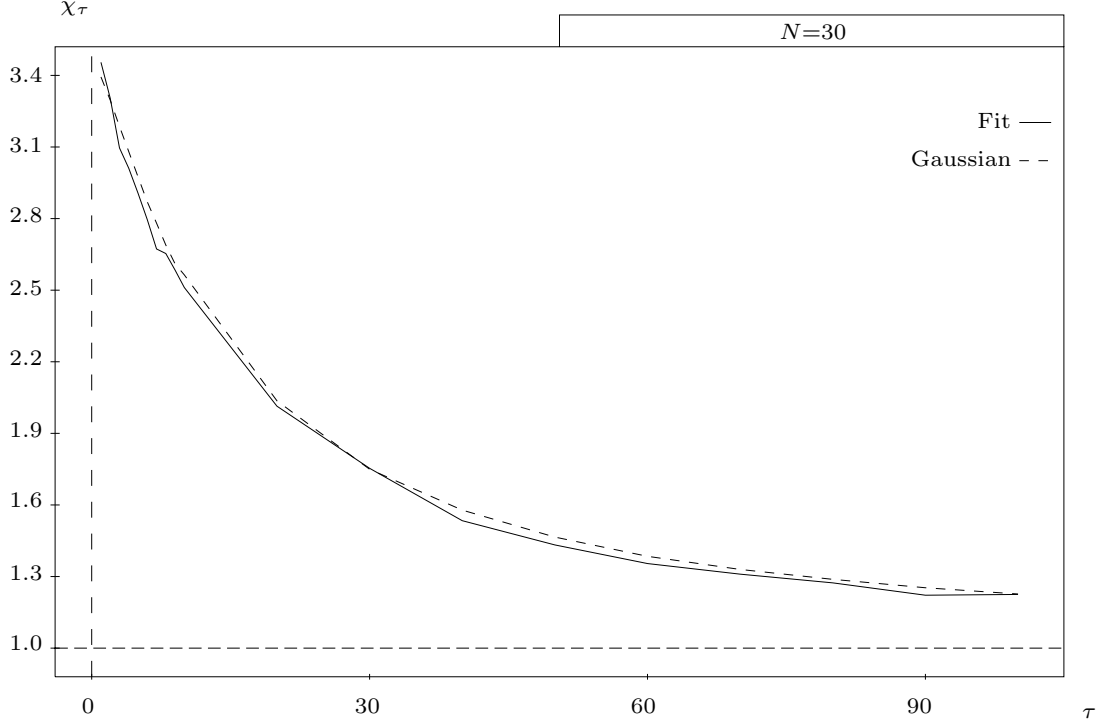
}{fig_ffp0}{Fig. \equ(fig5): Comparison between the value of
$\chi_\tau$ computed directly and using the Gaussian hypothesis for $N=30$. Similar result are found for $N=10,20,40$}{fig5}

The agreement is very good and justifies to use of eq.\equ(chi) for
large value of $\tau$.  The evaluated behavior of $\chi_\tau$ for
large value of $\tau$ from the Gaussian hypothesis is shown in
Fig. \equ(fig6).

\dimen2=25pt \advance\dimen2 by 126pt
\dimen3=210pt
\dimenfor{\dimen3}{166pt}
\eqfigfor{fig_fff}{
\ins{190pt}{0pt}{$\hbox to20pt{\hfill ${\scriptstyle \tau}$\hfill}$}
\ins{18pt}{\dimen2}{$\hbox to25pt{\hfill ${\scriptstyle \chi_\tau}$\hfill}$}
\ins{103pt}{144pt}{$\hbox to83pt{\hfill ${\scriptstyle N=10}$\hfill}$}
\ins{10pt}{3pt}{$\hbox to 38pt{\hfill${\scriptstyle 0}$\hfill}$}
\ins{49pt}{3pt}{$\hbox to 38pt{\hfill${\scriptstyle 2500}$\hfill}$}
\ins{87pt}{3pt}{$\hbox to 38pt{\hfill${\scriptstyle 5000}$\hfill}$}
\ins{125pt}{3pt}{$\hbox to 38pt{\hfill${\scriptstyle 7500}$\hfill}$}
\ins{163pt}{3pt}{$\hbox to 38pt{\hfill${\scriptstyle 10000}$\hfill}$}
\ins{0pt}{19pt}{$\hbox to 20pt{\hfill${\scriptstyle 0.95}$}$}
\ins{0pt}{42pt}{$\hbox to 20pt{\hfill${\scriptstyle 0.97}$}$}
\ins{0pt}{65pt}{$\hbox to 20pt{\hfill${\scriptstyle 0.99}$}$}
\ins{0pt}{88pt}{$\hbox to 20pt{\hfill${\scriptstyle 1.01}$}$}
\ins{0pt}{111pt}{$\hbox to 20pt{\hfill${\scriptstyle 1.03}$}$}
\ins{0pt}{134pt}{$\hbox to 20pt{\hfill${\scriptstyle 1.05}$}$}
}{\ins{190pt}{0pt}{$\hbox to20pt{\hfill ${\scriptstyle \tau}$\hfill}$}
\ins{18pt}{\dimen2}{$\hbox to25pt{\hfill ${\scriptstyle \chi_\tau}$\hfill}$}
\ins{103pt}{144pt}{$\hbox to83pt{\hfill ${\scriptstyle N=20}$\hfill}$}
\ins{10pt}{3pt}{$\hbox to 38pt{\hfill${\scriptstyle 0}$\hfill}$}
\ins{49pt}{3pt}{$\hbox to 38pt{\hfill${\scriptstyle 2500}$\hfill}$}
\ins{87pt}{3pt}{$\hbox to 38pt{\hfill${\scriptstyle 5000}$\hfill}$}
\ins{125pt}{3pt}{$\hbox to 38pt{\hfill${\scriptstyle 7500}$\hfill}$}
\ins{163pt}{3pt}{$\hbox to 38pt{\hfill${\scriptstyle 10000}$\hfill}$}
\ins{0pt}{19pt}{$\hbox to 20pt{\hfill${\scriptstyle 1.00}$}$}
\ins{0pt}{42pt}{$\hbox to 20pt{\hfill${\scriptstyle 1.02}$}$}
\ins{0pt}{65pt}{$\hbox to 20pt{\hfill${\scriptstyle 1.04}$}$}
\ins{0pt}{88pt}{$\hbox to 20pt{\hfill${\scriptstyle 1.06}$}$}
\ins{0pt}{111pt}{$\hbox to 20pt{\hfill${\scriptstyle 1.08}$}$}
\ins{0pt}{134pt}{$\hbox to 20pt{\hfill${\scriptstyle 1.10}$}$}
}{\ins{190pt}{0pt}{$\hbox to20pt{\hfill ${\scriptstyle \tau}$\hfill}$}
\ins{18pt}{\dimen2}{$\hbox to25pt{\hfill ${\scriptstyle \chi_\tau}$\hfill}$}
\ins{103pt}{144pt}{$\hbox to83pt{\hfill ${\scriptstyle N=30}$\hfill}$}
\ins{10pt}{3pt}{$\hbox to 38pt{\hfill${\scriptstyle 0}$\hfill}$}
\ins{49pt}{3pt}{$\hbox to 38pt{\hfill${\scriptstyle 2500}$\hfill}$}
\ins{87pt}{3pt}{$\hbox to 38pt{\hfill${\scriptstyle 5000}$\hfill}$}
\ins{125pt}{3pt}{$\hbox to 38pt{\hfill${\scriptstyle 7500}$\hfill}$}
\ins{163pt}{3pt}{$\hbox to 38pt{\hfill${\scriptstyle 10000}$\hfill}$}
\ins{0pt}{19pt}{$\hbox to 20pt{\hfill${\scriptstyle 1.0}$}$}
\ins{0pt}{58pt}{$\hbox to 20pt{\hfill${\scriptstyle 1.1}$}$}
\ins{0pt}{96pt}{$\hbox to 20pt{\hfill${\scriptstyle 1.2}$}$}
\ins{0pt}{134pt}{$\hbox to 20pt{\hfill${\scriptstyle 1.3}$}$}
}{\ins{190pt}{0pt}{$\hbox to20pt{\hfill ${\scriptstyle \tau}$\hfill}$}
\ins{18pt}{\dimen2}{$\hbox to25pt{\hfill ${\scriptstyle \chi_\tau}$\hfill}$}
\ins{103pt}{144pt}{$\hbox to83pt{\hfill ${\scriptstyle N=40}$\hfill}$}
\ins{10pt}{3pt}{$\hbox to 38pt{\hfill${\scriptstyle 0}$\hfill}$}
\ins{49pt}{3pt}{$\hbox to 38pt{\hfill${\scriptstyle 2500}$\hfill}$}
\ins{87pt}{3pt}{$\hbox to 38pt{\hfill${\scriptstyle 5000}$\hfill}$}
\ins{125pt}{3pt}{$\hbox to 38pt{\hfill${\scriptstyle 7500}$\hfill}$}
\ins{163pt}{3pt}{$\hbox to 38pt{\hfill${\scriptstyle 10000}$\hfill}$}
\ins{0pt}{19pt}{$\hbox to 20pt{\hfill${\scriptstyle 1.0}$}$}
\ins{0pt}{48pt}{$\hbox to 20pt{\hfill${\scriptstyle 1.1}$}$}
\ins{0pt}{77pt}{$\hbox to 20pt{\hfill${\scriptstyle 1.2}$}$}
\ins{0pt}{106pt}{$\hbox to 20pt{\hfill${\scriptstyle 1.3}$}$}
\ins{0pt}{134pt}{$\hbox to 20pt{\hfill${\scriptstyle 1.4}$}$}
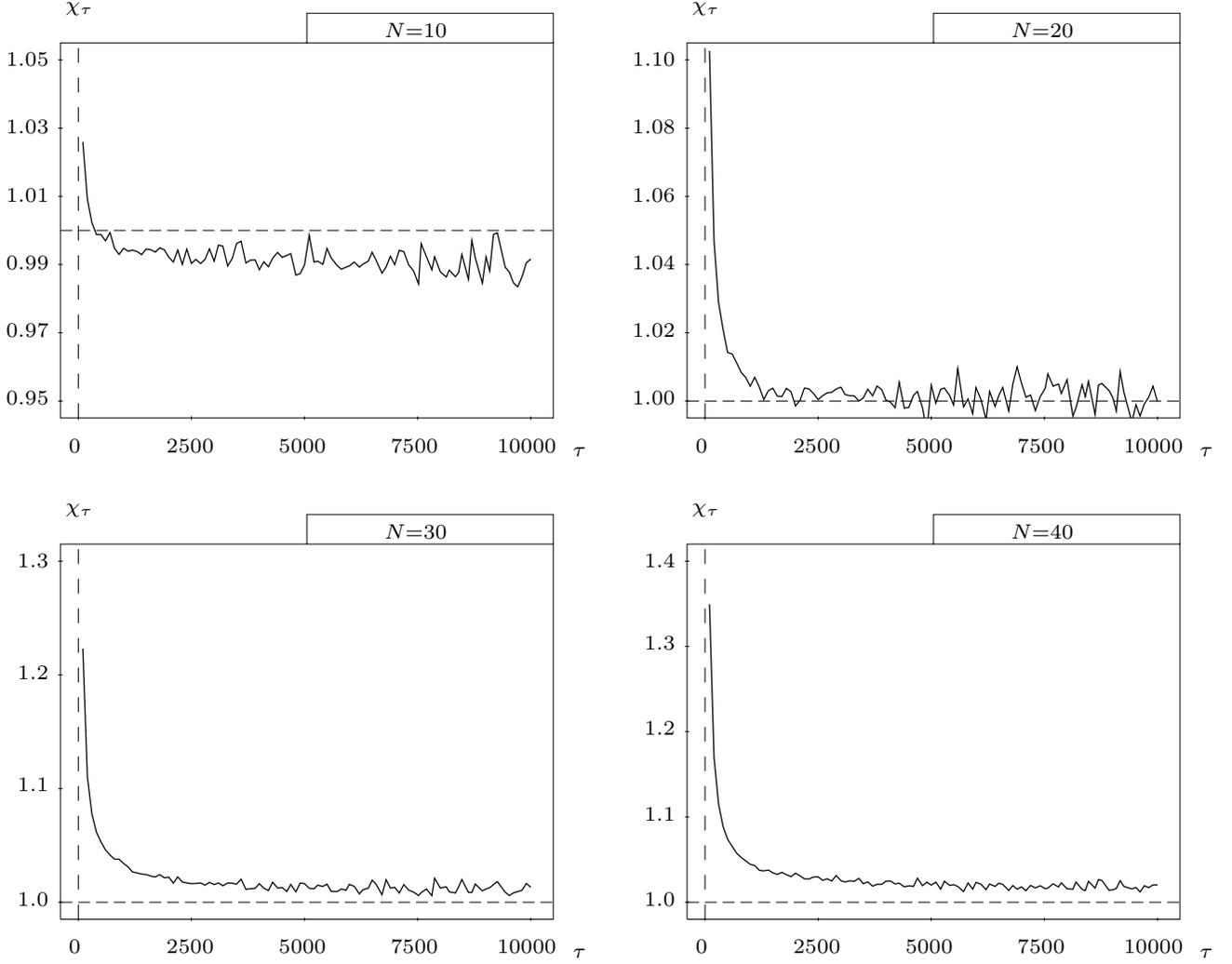
}{Fig. \equ(fig6): Value of $\chi_\tau$ obtained from the
Gaussian hypothesis for large $\tau$}{fig6}

In all cases the values of $\chi_\tau$ are very close to $1$ and we can
therefore say that the prediction of the fluctuation law appears to be
verified by our numerical simulations.

Finally observe that the approach to 1 of $\chi_\tau$ can be connected
to the decay property of the auto-correlation $D(t)=\langle
\sigma(S^t(\cdot))\sigma(\cdot)\rangle-1$. In fact we have:

$$C(\tau)=\frac{2}{\tau}\sum_{t=-\tau}^{\tau}D(t)-
\frac{2}{\tau^{2}}\sum_{t=-\tau}^{\tau}|t|D(t)\Eq(dec)$$

The fast approach of $\chi_\tau$ to its limit can thus be interpreted
as a rapid decay of the correlation for the phase space contraction of
the system. We will see that this behavior changes greatly when we
consider partial phase space contractions.

\newsec{Local fluctuations}

An interesting question is whether one can give a local version of the
fluctuation theorem. To this extent observe that we can write
$\alpha_\tau(X)=\alpha^\su_\tau(X)+ \alpha^\giu_\tau(X)$ where

$$\alpha^\su_\tau(X)=\sum_{i=-\tau/2}^{\tau/2}\alpha(S^i(X))\chi^\su(S^i(X))
\Eq(atausu)$$
and $\chi^\su(x)=1$ if $X$ correspond to a collision of a particle
with the upper wall and 0 otherwise. An analogous definition holds for
$\alpha^\giu_\tau(X)$.

The generalized version of the fluctuation theorem of \cita{G} gives

$$\lim_{\tau\to\infty}\frac{1}{\media{\alpha^\su_\tau}_+}
\log\frac{\pi^{\su\giu}_\tau(p_1,p_2)}{\pi^{\su\giu}_\tau(-p_1,-p_2)}=
p_1+p_2\Eq(gflu)$$
where $\pi^{\su\giu}_\tau(p_1,p_2)$ is the joint distribution of
$\sigma^\su_\tau(X)=\alpha^\su_\tau(X)/\media{\alpha^\su_\tau}_+$ and
$\sigma^\giu_\tau(X)=\alpha^\giu_\tau(X)/\media{\alpha^\giu_\tau}_+$.
Due to the symmetry of the problem the two variables
$\sigma^\su_\tau(X)$ and $\sigma^\giu_\tau(X)$ can be assumed to be
identically distributed\annota{1}{Although this is well verified
numerically it is not evident form a theoretical point of view. It is
in fact easy to see that if the particles do not interact, \ie they do
not collide, the mean momentum of the center of mass is, generically,
not zero and this will probably create asymmetry between the two
walls. This phenomenon is probably destroyed by the ``mixing''
behavior generated by the binary collisions. It must be nevertheless
noted that for small systems, like ours, the fluctuations of the
center of mass momentum are quite big and its correlations decay
slowly}.  It is easy to see that if we suppose them independent, both
of them will separately satisfy the fluctuation theorem:

$$\lim_{\tau\to\infty}x_\tau^\su(p)=p\Eq(lflu_s)$$
where

$$x_\tau^\su(p)=\frac{1}{\media{\alpha^\su_\tau}_+}
\log\frac{\pi^{\su}_\tau(p)}{\pi^{\su}_\tau(-p)}\Eq(flu_s)$$
This can be checked as we did for $\alpha_\tau(x)$. In Fig. \equ(fig7)
we show the graph of $x^\su_\tau(p)$ for $N=20$ and several values of $\tau$.

\dimen2=25pt \advance\dimen2 by 158pt
\dimen3=196pt
\dimenfor{\dimen3}{199pt}
\eqfigfor{fig_su}{
\ins{176pt}{0pt}{$\hbox to20pt{\hfill ${\scriptstyle p}$\hfill}$}
\ins{12pt}{\dimen2}{$\hbox to25pt{\hfill ${\scriptstyle x^\su(p)}$\hfill}$}
\ins{99pt}{176pt}{$\hbox to79pt{\hfill ${\scriptstyle \tau=100}$\hfill}$}
\ins{7pt}{3pt}{$\hbox to 35pt{\hfill${\scriptstyle 0}$\hfill}$}
\ins{42pt}{3pt}{$\hbox to 35pt{\hfill${\scriptstyle 1}$\hfill}$}
\ins{77pt}{3pt}{$\hbox to 35pt{\hfill${\scriptstyle 2}$\hfill}$}
\ins{113pt}{3pt}{$\hbox to 35pt{\hfill${\scriptstyle 3}$\hfill}$}
\ins{148pt}{3pt}{$\hbox to 35pt{\hfill${\scriptstyle 4}$\hfill}$}
\ins{0pt}{19pt}{$\hbox to 14pt{\hfill${\scriptstyle 0}$}$}
\ins{0pt}{38pt}{$\hbox to 14pt{\hfill${\scriptstyle 1}$}$}
\ins{0pt}{57pt}{$\hbox to 14pt{\hfill${\scriptstyle 2}$}$}
\ins{0pt}{76pt}{$\hbox to 14pt{\hfill${\scriptstyle 3}$}$}
\ins{0pt}{94pt}{$\hbox to 14pt{\hfill${\scriptstyle 4}$}$}
\ins{0pt}{113pt}{$\hbox to 14pt{\hfill${\scriptstyle 5}$}$}
\ins{0pt}{132pt}{$\hbox to 14pt{\hfill${\scriptstyle 6}$}$}
\ins{0pt}{150pt}{$\hbox to 14pt{\hfill${\scriptstyle 7}$}$}
\ins{0pt}{169pt}{$\hbox to 14pt{\hfill${\scriptstyle 8}$}$}
}{\ins{176pt}{0pt}{$\hbox to20pt{\hfill ${\scriptstyle p}$\hfill}$}
\ins{12pt}{\dimen2}{$\hbox to25pt{\hfill ${\scriptstyle x^\su(p)}$\hfill}$}
\ins{99pt}{176pt}{$\hbox to79pt{\hfill ${\scriptstyle \tau=800}$\hfill}$}
\ins{-3pt}{3pt}{$\hbox to 55pt{\hfill${\scriptstyle 0.0}$\hfill}$}
\ins{52pt}{3pt}{$\hbox to 55pt{\hfill${\scriptstyle 0.5}$\hfill}$}
\ins{108pt}{3pt}{$\hbox to 55pt{\hfill${\scriptstyle 1.0}$\hfill}$}
\ins{0pt}{17pt}{$\hbox to 14pt{\hfill${\scriptstyle 0.0}$}$}
\ins{0pt}{59pt}{$\hbox to 14pt{\hfill${\scriptstyle 0.5}$}$}
\ins{0pt}{101pt}{$\hbox to 14pt{\hfill${\scriptstyle 1.0}$}$}
\ins{0pt}{143pt}{$\hbox to 14pt{\hfill${\scriptstyle 1.5}$}$}
}{\ins{176pt}{0pt}{$\hbox to20pt{\hfill ${\scriptstyle p}$\hfill}$}
\ins{12pt}{\dimen2}{$\hbox to25pt{\hfill ${\scriptstyle x^\su(p)}$\hfill}$}
\ins{99pt}{176pt}{$\hbox to79pt{\hfill ${\scriptstyle \tau=1400}$\hfill}$}
\ins{1pt}{3pt}{$\hbox to 48pt{\hfill${\scriptstyle 0.0}$\hfill}$}
\ins{49pt}{3pt}{$\hbox to 48pt{\hfill${\scriptstyle 0.3}$\hfill}$}
\ins{97pt}{3pt}{$\hbox to 48pt{\hfill${\scriptstyle 0.6}$\hfill}$}
\ins{145pt}{3pt}{$\hbox to 48pt{\hfill${\scriptstyle 0.9}$\hfill}$}
\ins{0pt}{14pt}{$\hbox to 14pt{\hfill${\scriptstyle 0.0}$}$}
\ins{0pt}{69pt}{$\hbox to 14pt{\hfill${\scriptstyle 0.3}$}$}
\ins{0pt}{125pt}{$\hbox to 14pt{\hfill${\scriptstyle 0.6}$}$}
}{\ins{176pt}{0pt}{$\hbox to20pt{\hfill ${\scriptstyle p}$\hfill}$}
\ins{12pt}{\dimen2}{$\hbox to25pt{\hfill ${\scriptstyle x^\su(p)}$\hfill}$}
\ins{99pt}{176pt}{$\hbox to79pt{\hfill ${\scriptstyle \tau=2000}$\hfill}$}
\ins{7pt}{3pt}{$\hbox to 36pt{\hfill${\scriptstyle 0.0}$\hfill}$}
\ins{43pt}{3pt}{$\hbox to 36pt{\hfill${\scriptstyle 0.2}$\hfill}$}
\ins{79pt}{3pt}{$\hbox to 36pt{\hfill${\scriptstyle 0.4}$\hfill}$}
\ins{115pt}{3pt}{$\hbox to 36pt{\hfill${\scriptstyle 0.6}$\hfill}$}
\ins{151pt}{3pt}{$\hbox to 36pt{\hfill${\scriptstyle 0.8}$\hfill}$}
\ins{0pt}{15pt}{$\hbox to 14pt{\hfill${\scriptstyle 0.0}$}$}
\ins{0pt}{56pt}{$\hbox to 14pt{\hfill${\scriptstyle 0.2}$}$}
\ins{0pt}{97pt}{$\hbox to 14pt{\hfill${\scriptstyle 0.4}$}$}
\ins{0pt}{137pt}{$\hbox to 14pt{\hfill${\scriptstyle 0.6}$}$}
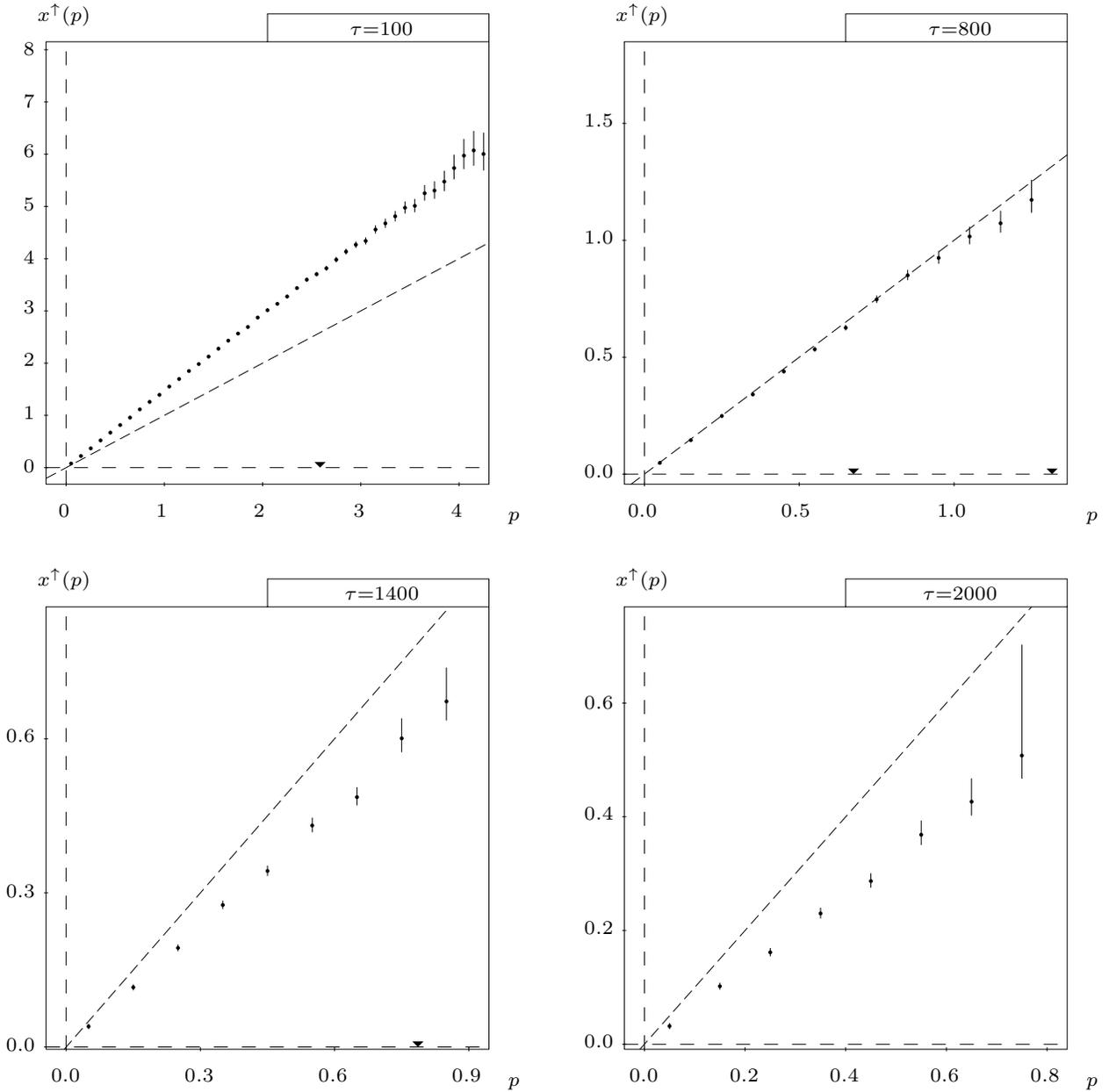
}{Fig. \equ(fig7): Graphs of $x^\su_\tau(p)$ for
$\tau=100,800,1400,2000$ for $N=20$. The arrows represents the value 1 
plus o minus the standard deviation of $\sigma^\su_\tau$.}{fig7}

Observe that here we are able to construct the distribution function
$\pi_\tau$ up to $\tau=2000$. This is not surprising considering that
$\alpha^\su_\tau(X)$ is, in the mean, the sum of $\tau/2$ non zero
terms so that it can be expected to have fluctuation roughly similar to
those of $\alpha_{\tau/2}(X)$.

As before we can look at the best fit of $x_\tau^\su(p)$ of the form
$x_\tau^\su(p)=\chi^\su_\tau p$. The results are shown in
Fig. \equ(fig8).

\dimen2=25pt \advance\dimen2 by 126pt
\dimen3=204pt
\dimenfor{\dimen3}{166pt}
\eqfigfor{fig_fs}{
\ins{184pt}{0pt}{$\hbox to20pt{\hfill ${\scriptstyle \tau}$\hfill}$}
\ins{12pt}{\dimen2}{$\hbox to25pt{\hfill ${\scriptstyle \chi^\su_\tau}$\hfill}$}
\ins{103pt}{144pt}{$\hbox to83pt{\hfill ${\scriptstyle N=10}$\hfill}$}
\ins{10pt}{3pt}{$\hbox to 30pt{\hfill${\scriptstyle 0}$\hfill}$}
\ins{40pt}{3pt}{$\hbox to 30pt{\hfill${\scriptstyle 400}$\hfill}$}
\ins{70pt}{3pt}{$\hbox to 30pt{\hfill${\scriptstyle 800}$\hfill}$}
\ins{101pt}{3pt}{$\hbox to 30pt{\hfill${\scriptstyle 1200}$\hfill}$}
\ins{131pt}{3pt}{$\hbox to 30pt{\hfill${\scriptstyle 1600}$\hfill}$}
\ins{161pt}{3pt}{$\hbox to 30pt{\hfill${\scriptstyle 2000}$\hfill}$}
\ins{0pt}{21pt}{$\hbox to 14pt{\hfill${\scriptstyle 0.3}$}$}
\ins{0pt}{37pt}{$\hbox to 14pt{\hfill${\scriptstyle 0.6}$}$}
\ins{0pt}{53pt}{$\hbox to 14pt{\hfill${\scriptstyle 0.9}$}$}
\ins{0pt}{70pt}{$\hbox to 14pt{\hfill${\scriptstyle 1.2}$}$}
\ins{0pt}{86pt}{$\hbox to 14pt{\hfill${\scriptstyle 1.5}$}$}
\ins{0pt}{102pt}{$\hbox to 14pt{\hfill${\scriptstyle 1.8}$}$}
\ins{0pt}{119pt}{$\hbox to 14pt{\hfill${\scriptstyle 2.1}$}$}
\ins{0pt}{135pt}{$\hbox to 14pt{\hfill${\scriptstyle 2.4}$}$}
}{\ins{184pt}{0pt}{$\hbox to20pt{\hfill ${\scriptstyle \tau}$\hfill}$}
\ins{12pt}{\dimen2}{$\hbox to25pt{\hfill ${\scriptstyle \chi^\su_\tau}$\hfill}$}
\ins{103pt}{144pt}{$\hbox to83pt{\hfill ${\scriptstyle N=20}$\hfill}$}
\ins{10pt}{3pt}{$\hbox to 30pt{\hfill${\scriptstyle 0}$\hfill}$}
\ins{40pt}{3pt}{$\hbox to 30pt{\hfill${\scriptstyle 400}$\hfill}$}
\ins{70pt}{3pt}{$\hbox to 30pt{\hfill${\scriptstyle 800}$\hfill}$}
\ins{101pt}{3pt}{$\hbox to 30pt{\hfill${\scriptstyle 1200}$\hfill}$}
\ins{131pt}{3pt}{$\hbox to 30pt{\hfill${\scriptstyle 1600}$\hfill}$}
\ins{161pt}{3pt}{$\hbox to 30pt{\hfill${\scriptstyle 2000}$\hfill}$}
\ins{0pt}{20pt}{$\hbox to 14pt{\hfill${\scriptstyle 0.6}$}$}
\ins{0pt}{33pt}{$\hbox to 14pt{\hfill${\scriptstyle 0.9}$}$}
\ins{0pt}{47pt}{$\hbox to 14pt{\hfill${\scriptstyle 1.2}$}$}
\ins{0pt}{61pt}{$\hbox to 14pt{\hfill${\scriptstyle 1.5}$}$}
\ins{0pt}{75pt}{$\hbox to 14pt{\hfill${\scriptstyle 1.8}$}$}
\ins{0pt}{88pt}{$\hbox to 14pt{\hfill${\scriptstyle 2.1}$}$}
\ins{0pt}{102pt}{$\hbox to 14pt{\hfill${\scriptstyle 2.4}$}$}
\ins{0pt}{116pt}{$\hbox to 14pt{\hfill${\scriptstyle 2.7}$}$}
\ins{0pt}{130pt}{$\hbox to 14pt{\hfill${\scriptstyle 3.0}$}$}
}{\ins{184pt}{0pt}{$\hbox to20pt{\hfill ${\scriptstyle \tau}$\hfill}$}
\ins{12pt}{\dimen2}{$\hbox to25pt{\hfill ${\scriptstyle \chi^\su_\tau}$\hfill}$}
\ins{103pt}{144pt}{$\hbox to83pt{\hfill ${\scriptstyle N=30}$\hfill}$}
\ins{10pt}{3pt}{$\hbox to 30pt{\hfill${\scriptstyle 0}$\hfill}$}
\ins{40pt}{3pt}{$\hbox to 30pt{\hfill${\scriptstyle 400}$\hfill}$}
\ins{70pt}{3pt}{$\hbox to 30pt{\hfill${\scriptstyle 800}$\hfill}$}
\ins{101pt}{3pt}{$\hbox to 30pt{\hfill${\scriptstyle 1200}$\hfill}$}
\ins{131pt}{3pt}{$\hbox to 30pt{\hfill${\scriptstyle 1600}$\hfill}$}
\ins{161pt}{3pt}{$\hbox to 30pt{\hfill${\scriptstyle 2000}$\hfill}$}
\ins{0pt}{21pt}{$\hbox to 14pt{\hfill${\scriptstyle 0.8}$}$}
\ins{0pt}{34pt}{$\hbox to 14pt{\hfill${\scriptstyle 1.1}$}$}
\ins{0pt}{47pt}{$\hbox to 14pt{\hfill${\scriptstyle 1.4}$}$}
\ins{0pt}{60pt}{$\hbox to 14pt{\hfill${\scriptstyle 1.7}$}$}
\ins{0pt}{72pt}{$\hbox to 14pt{\hfill${\scriptstyle 2.0}$}$}
\ins{0pt}{85pt}{$\hbox to 14pt{\hfill${\scriptstyle 2.3}$}$}
\ins{0pt}{98pt}{$\hbox to 14pt{\hfill${\scriptstyle 2.6}$}$}
\ins{0pt}{111pt}{$\hbox to 14pt{\hfill${\scriptstyle 2.9}$}$}
\ins{0pt}{124pt}{$\hbox to 14pt{\hfill${\scriptstyle 3.2}$}$}
\ins{0pt}{137pt}{$\hbox to 14pt{\hfill${\scriptstyle 3.5}$}$}
}{\ins{184pt}{0pt}{$\hbox to20pt{\hfill ${\scriptstyle \tau}$\hfill}$}
\ins{12pt}{\dimen2}{$\hbox to25pt{\hfill ${\scriptstyle \chi^\su_\tau}$\hfill}$}
\ins{103pt}{144pt}{$\hbox to83pt{\hfill ${\scriptstyle N=40}$\hfill}$}
\ins{10pt}{3pt}{$\hbox to 30pt{\hfill${\scriptstyle 0}$\hfill}$}
\ins{40pt}{3pt}{$\hbox to 30pt{\hfill${\scriptstyle 300}$\hfill}$}
\ins{70pt}{3pt}{$\hbox to 30pt{\hfill${\scriptstyle 600}$\hfill}$}
\ins{101pt}{3pt}{$\hbox to 30pt{\hfill${\scriptstyle 900}$\hfill}$}
\ins{131pt}{3pt}{$\hbox to 30pt{\hfill${\scriptstyle 1200}$\hfill}$}
\ins{161pt}{3pt}{$\hbox to 30pt{\hfill${\scriptstyle 1500}$\hfill}$}
\ins{0pt}{15pt}{$\hbox to 14pt{\hfill${\scriptstyle 0.9}$}$}
\ins{0pt}{27pt}{$\hbox to 14pt{\hfill${\scriptstyle 1.2}$}$}
\ins{0pt}{39pt}{$\hbox to 14pt{\hfill${\scriptstyle 1.5}$}$}
\ins{0pt}{50pt}{$\hbox to 14pt{\hfill${\scriptstyle 1.8}$}$}
\ins{0pt}{62pt}{$\hbox to 14pt{\hfill${\scriptstyle 2.1}$}$}
\ins{0pt}{74pt}{$\hbox to 14pt{\hfill${\scriptstyle 2.4}$}$}
\ins{0pt}{86pt}{$\hbox to 14pt{\hfill${\scriptstyle 2.7}$}$}
\ins{0pt}{97pt}{$\hbox to 14pt{\hfill${\scriptstyle 3.0}$}$}
\ins{0pt}{109pt}{$\hbox to 14pt{\hfill${\scriptstyle 3.3}$}$}
\ins{0pt}{121pt}{$\hbox to 14pt{\hfill${\scriptstyle 3.6}$}$}
\ins{0pt}{132pt}{$\hbox to 14pt{\hfill${\scriptstyle 3.9}$}$}
}{Fig. \equ(fig8): Behavior of $\chi_\tau^\su$ as a 
function of $\tau$ for $N=10,20,30$ and $40$.}{fig8}

It is clear that for $N=10$ and $20$ the fit goes significatively
below 1 at a value of $\tau$ for which $\chi_\tau$ is very near its
theoretical value of 1. Moreover $\chi^\su_\tau$ does not seem to have
reached a limiting value while $\chi_\tau$ reaches its limiting value of
one quite soon.

The cases $N=30$ and $40$ are less clear because we are unable to
construct the distribution function for $\tau$ large enough to have a
clear idea of the limiting value of $\chi_\tau$. But it seems
reasonable to deduce from the graph that also in this case $\chi_\tau$
will became smaller that 1.

Considering the argument at the beginning of this section it is
interesting to look at the cross correlation between $\sigma^\su_\tau$
and $\sigma^\giu_\tau$ given by:

$$C^{\su\giu}_\tau=\langle \sigma^\su_\tau
\sigma^\giu_\tau\rangle_+-1$$

whose graph is plotted in Fig. \equ(fig9).

\dimen2=25pt \advance\dimen2 by 110pt
\dimen3=210pt
\dimenfor{\dimen3}{149pt}
\eqfigfor{fig_sg}{
\ins{190pt}{0pt}{$\hbox to20pt{\hfill ${\scriptstyle \tau}$\hfill}$}
\ins{18pt}{\dimen2}{$\hbox to25pt{\hfill ${\scriptstyle C^{\su\giu}_\tau}$\hfill}$}
\ins{103pt}{128pt}{$\hbox to83pt{\hfill ${\scriptstyle N=10}$\hfill}$}
\ins{16pt}{3pt}{$\hbox to 30pt{\hfill${\scriptstyle 0}$\hfill}$}
\ins{46pt}{3pt}{$\hbox to 30pt{\hfill${\scriptstyle 400}$\hfill}$}
\ins{76pt}{3pt}{$\hbox to 30pt{\hfill${\scriptstyle 800}$\hfill}$}
\ins{107pt}{3pt}{$\hbox to 30pt{\hfill${\scriptstyle 1200}$\hfill}$}
\ins{137pt}{3pt}{$\hbox to 30pt{\hfill${\scriptstyle 1600}$\hfill}$}
\ins{167pt}{3pt}{$\hbox to 30pt{\hfill${\scriptstyle 2000}$\hfill}$}
\ins{0pt}{16pt}{$\hbox to 20pt{\hfill${\scriptstyle -0.6}$}$}
\ins{0pt}{27pt}{$\hbox to 20pt{\hfill${\scriptstyle 0.0}$}$}
\ins{0pt}{37pt}{$\hbox to 20pt{\hfill${\scriptstyle 0.6}$}$}
\ins{0pt}{48pt}{$\hbox to 20pt{\hfill${\scriptstyle 1.2}$}$}
\ins{0pt}{59pt}{$\hbox to 20pt{\hfill${\scriptstyle 1.8}$}$}
\ins{0pt}{69pt}{$\hbox to 20pt{\hfill${\scriptstyle 2.4}$}$}
\ins{0pt}{80pt}{$\hbox to 20pt{\hfill${\scriptstyle 3.0}$}$}
\ins{0pt}{91pt}{$\hbox to 20pt{\hfill${\scriptstyle 3.6}$}$}
\ins{0pt}{102pt}{$\hbox to 20pt{\hfill${\scriptstyle 4.2}$}$}
\ins{0pt}{112pt}{$\hbox to 20pt{\hfill${\scriptstyle 4.8}$}$}
\ins{0pt}{123pt}{$\hbox to 20pt{\hfill${\scriptstyle 5.4}$}$}
}{\ins{190pt}{0pt}{$\hbox to20pt{\hfill ${\scriptstyle \tau}$\hfill}$}
\ins{18pt}{\dimen2}{$\hbox to25pt{\hfill ${\scriptstyle C^{\su\giu}_\tau}$\hfill}$}
\ins{103pt}{128pt}{$\hbox to83pt{\hfill ${\scriptstyle N=20}$\hfill}$}
\ins{16pt}{3pt}{$\hbox to 30pt{\hfill${\scriptstyle 0}$\hfill}$}
\ins{46pt}{3pt}{$\hbox to 30pt{\hfill${\scriptstyle 400}$\hfill}$}
\ins{76pt}{3pt}{$\hbox to 30pt{\hfill${\scriptstyle 800}$\hfill}$}
\ins{107pt}{3pt}{$\hbox to 30pt{\hfill${\scriptstyle 1200}$\hfill}$}
\ins{137pt}{3pt}{$\hbox to 30pt{\hfill${\scriptstyle 1600}$\hfill}$}
\ins{167pt}{3pt}{$\hbox to 30pt{\hfill${\scriptstyle 2000}$\hfill}$}
\ins{0pt}{21pt}{$\hbox to 20pt{\hfill${\scriptstyle 0.0}$}$}
\ins{0pt}{33pt}{$\hbox to 20pt{\hfill${\scriptstyle 0.2}$}$}
\ins{0pt}{45pt}{$\hbox to 20pt{\hfill${\scriptstyle 0.4}$}$}
\ins{0pt}{56pt}{$\hbox to 20pt{\hfill${\scriptstyle 0.6}$}$}
\ins{0pt}{68pt}{$\hbox to 20pt{\hfill${\scriptstyle 0.8}$}$}
\ins{0pt}{80pt}{$\hbox to 20pt{\hfill${\scriptstyle 1.0}$}$}
\ins{0pt}{91pt}{$\hbox to 20pt{\hfill${\scriptstyle 1.2}$}$}
\ins{0pt}{103pt}{$\hbox to 20pt{\hfill${\scriptstyle 1.4}$}$}
\ins{0pt}{115pt}{$\hbox to 20pt{\hfill${\scriptstyle 1.6}$}$}
}{\ins{190pt}{0pt}{$\hbox to20pt{\hfill ${\scriptstyle \tau}$\hfill}$}
\ins{18pt}{\dimen2}{$\hbox to25pt{\hfill ${\scriptstyle C^{\su\giu}_\tau}$\hfill}$}
\ins{103pt}{128pt}{$\hbox to83pt{\hfill ${\scriptstyle N=30}$\hfill}$}
\ins{16pt}{3pt}{$\hbox to 30pt{\hfill${\scriptstyle 0}$\hfill}$}
\ins{46pt}{3pt}{$\hbox to 30pt{\hfill${\scriptstyle 400}$\hfill}$}
\ins{76pt}{3pt}{$\hbox to 30pt{\hfill${\scriptstyle 800}$\hfill}$}
\ins{107pt}{3pt}{$\hbox to 30pt{\hfill${\scriptstyle 1200}$\hfill}$}
\ins{137pt}{3pt}{$\hbox to 30pt{\hfill${\scriptstyle 1600}$\hfill}$}
\ins{167pt}{3pt}{$\hbox to 30pt{\hfill${\scriptstyle 2000}$\hfill}$}
\ins{0pt}{19pt}{$\hbox to 20pt{\hfill${\scriptstyle -1.0}$}$}
\ins{0pt}{35pt}{$\hbox to 20pt{\hfill${\scriptstyle -0.7}$}$}
\ins{0pt}{52pt}{$\hbox to 20pt{\hfill${\scriptstyle -0.4}$}$}
\ins{0pt}{69pt}{$\hbox to 20pt{\hfill${\scriptstyle -0.1}$}$}
\ins{0pt}{86pt}{$\hbox to 20pt{\hfill${\scriptstyle 0.2}$}$}
\ins{0pt}{103pt}{$\hbox to 20pt{\hfill${\scriptstyle 0.5}$}$}
\ins{0pt}{119pt}{$\hbox to 20pt{\hfill${\scriptstyle 0.8}$}$}
}{\ins{190pt}{0pt}{$\hbox to20pt{\hfill ${\scriptstyle \tau}$\hfill}$}
\ins{18pt}{\dimen2}{$\hbox to25pt{\hfill ${\scriptstyle C^{\su\giu}_\tau}$\hfill}$}
\ins{103pt}{128pt}{$\hbox to83pt{\hfill ${\scriptstyle N=40}$\hfill}$}
\ins{16pt}{3pt}{$\hbox to 30pt{\hfill${\scriptstyle 0}$\hfill}$}
\ins{46pt}{3pt}{$\hbox to 30pt{\hfill${\scriptstyle 400}$\hfill}$}
\ins{76pt}{3pt}{$\hbox to 30pt{\hfill${\scriptstyle 800}$\hfill}$}
\ins{107pt}{3pt}{$\hbox to 30pt{\hfill${\scriptstyle 1200}$\hfill}$}
\ins{137pt}{3pt}{$\hbox to 30pt{\hfill${\scriptstyle 1600}$\hfill}$}
\ins{167pt}{3pt}{$\hbox to 30pt{\hfill${\scriptstyle 2000}$\hfill}$}
\ins{0pt}{19pt}{$\hbox to 20pt{\hfill${\scriptstyle -1.0}$}$}
\ins{0pt}{32pt}{$\hbox to 20pt{\hfill${\scriptstyle -0.8}$}$}
\ins{0pt}{45pt}{$\hbox to 20pt{\hfill${\scriptstyle -0.6}$}$}
\ins{0pt}{59pt}{$\hbox to 20pt{\hfill${\scriptstyle -0.4}$}$}
\ins{0pt}{72pt}{$\hbox to 20pt{\hfill${\scriptstyle -0.2}$}$}
\ins{0pt}{86pt}{$\hbox to 20pt{\hfill${\scriptstyle -0.0}$}$}
\ins{0pt}{99pt}{$\hbox to 20pt{\hfill${\scriptstyle 0.2}$}$}
\ins{0pt}{113pt}{$\hbox to 20pt{\hfill${\scriptstyle 0.4}$}$}
}{Fig. \equ(fig9): Behavior of $C_\tau^{\su\giu}$ as a 
function of $\tau$ for $N=10,20,30$ and $40$.}{fig9}

It is interesting to note that, if we disregard the value of
$C^{\su\giu}_\tau$ for small value of $\tau$, we have behavior of
$C^{\su\giu}_\tau$ is linked to that of $\chi_\tau$. In
particular we can see that when $C^{\su\giu}_\tau=0$ we have
$\chi_\tau=1$. This is particularly clear in the case $N=10$ and $20$.

This suggests that the higher order correlations of $\sigma^\su$
and  $\sigma^\giu$ are small compared to the
self-correlation. Moreover the distribution function $\pi^\su_\tau$
is again very well approximated by a Gaussian as can be seen in
Fig. \equ(fig10). 

\dimen2=25pt \advance\dimen2 by 158pt
\dimen3=202pt
\dimenfor{\dimen3}{199pt}
\eqfigfor{fig_gs}{
\ins{182pt}{0pt}{$\hbox to20pt{\hfill ${\scriptstyle p}$\hfill}$}
\ins{18pt}{\dimen2}{$\hbox to25pt{\hfill ${\scriptstyle \pi^\su_\tau}$\hfill}$}
\ins{99pt}{176pt}{$\hbox to79pt{\hfill ${\scriptstyle \tau=5}$\hfill}$}
\ins{0pt}{3pt}{$\hbox to 27pt{\hfill${\scriptstyle -18}$\hfill}$}
\ins{27pt}{3pt}{$\hbox to 27pt{\hfill${\scriptstyle -12}$\hfill}$}
\ins{54pt}{3pt}{$\hbox to 27pt{\hfill${\scriptstyle -6}$\hfill}$}
\ins{81pt}{3pt}{$\hbox to 27pt{\hfill${\scriptstyle 0}$\hfill}$}
\ins{108pt}{3pt}{$\hbox to 27pt{\hfill${\scriptstyle 6}$\hfill}$}
\ins{135pt}{3pt}{$\hbox to 27pt{\hfill${\scriptstyle 12}$\hfill}$}
\ins{162pt}{3pt}{$\hbox to 27pt{\hfill${\scriptstyle 18}$\hfill}$}
\ins{0pt}{21pt}{$\hbox to 20pt{\hfill${\scriptstyle 0.00}$}$}
\ins{0pt}{53pt}{$\hbox to 20pt{\hfill${\scriptstyle 0.03}$}$}
\ins{0pt}{86pt}{$\hbox to 20pt{\hfill${\scriptstyle 0.06}$}$}
\ins{0pt}{119pt}{$\hbox to 20pt{\hfill${\scriptstyle 0.09}$}$}
\ins{0pt}{152pt}{$\hbox to 20pt{\hfill${\scriptstyle 0.12}$}$}
}{\ins{182pt}{0pt}{$\hbox to20pt{\hfill ${\scriptstyle p}$\hfill}$}
\ins{18pt}{\dimen2}{$\hbox to25pt{\hfill ${\scriptstyle \pi^\su_\tau}$\hfill}$}
\ins{99pt}{176pt}{$\hbox to79pt{\hfill ${\scriptstyle \tau=50}$\hfill}$}
\ins{-1pt}{3pt}{$\hbox to 36pt{\hfill${\scriptstyle -6}$\hfill}$}
\ins{36pt}{3pt}{$\hbox to 36pt{\hfill${\scriptstyle -3}$\hfill}$}
\ins{73pt}{3pt}{$\hbox to 36pt{\hfill${\scriptstyle 0}$\hfill}$}
\ins{110pt}{3pt}{$\hbox to 36pt{\hfill${\scriptstyle 3}$\hfill}$}
\ins{147pt}{3pt}{$\hbox to 36pt{\hfill${\scriptstyle 6}$\hfill}$}
\ins{0pt}{21pt}{$\hbox to 20pt{\hfill${\scriptstyle 0.0}$}$}
\ins{0pt}{66pt}{$\hbox to 20pt{\hfill${\scriptstyle 0.1}$}$}
\ins{0pt}{112pt}{$\hbox to 20pt{\hfill${\scriptstyle 0.2}$}$}
\ins{0pt}{157pt}{$\hbox to 20pt{\hfill${\scriptstyle 0.3}$}$}
}{\ins{182pt}{0pt}{$\hbox to20pt{\hfill ${\scriptstyle p}$\hfill}$}
\ins{18pt}{\dimen2}{$\hbox to25pt{\hfill ${\scriptstyle \pi^\su_\tau}$\hfill}$}
\ins{99pt}{176pt}{$\hbox to79pt{\hfill ${\scriptstyle \tau=500}$\hfill}$}
\ins{0pt}{3pt}{$\hbox to 29pt{\hfill${\scriptstyle -2}$\hfill}$}
\ins{29pt}{3pt}{$\hbox to 29pt{\hfill${\scriptstyle -1}$\hfill}$}
\ins{59pt}{3pt}{$\hbox to 29pt{\hfill${\scriptstyle 0}$\hfill}$}
\ins{88pt}{3pt}{$\hbox to 29pt{\hfill${\scriptstyle 1}$\hfill}$}
\ins{117pt}{3pt}{$\hbox to 29pt{\hfill${\scriptstyle 2}$\hfill}$}
\ins{147pt}{3pt}{$\hbox to 29pt{\hfill${\scriptstyle 3}$\hfill}$}
\ins{0pt}{21pt}{$\hbox to 20pt{\hfill${\scriptstyle 0.0}$}$}
\ins{0pt}{57pt}{$\hbox to 20pt{\hfill${\scriptstyle 0.2}$}$}
\ins{0pt}{93pt}{$\hbox to 20pt{\hfill${\scriptstyle 0.4}$}$}
\ins{0pt}{129pt}{$\hbox to 20pt{\hfill${\scriptstyle 0.6}$}$}
\ins{0pt}{165pt}{$\hbox to 20pt{\hfill${\scriptstyle 0.8}$}$}
}{\ins{182pt}{0pt}{$\hbox to20pt{\hfill ${\scriptstyle p}$\hfill}$}
\ins{18pt}{\dimen2}{$\hbox to25pt{\hfill ${\scriptstyle \pi^\su_\tau}$\hfill}$}
\ins{99pt}{176pt}{$\hbox to79pt{\hfill ${\scriptstyle \tau=1000}$\hfill}$}
\ins{32pt}{3pt}{$\hbox to 28pt{\hfill${\scriptstyle -0.8}$\hfill}$}
\ins{61pt}{3pt}{$\hbox to 28pt{\hfill${\scriptstyle 0.0}$\hfill}$}
\ins{89pt}{3pt}{$\hbox to 28pt{\hfill${\scriptstyle 0.8}$\hfill}$}
\ins{117pt}{3pt}{$\hbox to 28pt{\hfill${\scriptstyle 1.6}$\hfill}$}
\ins{145pt}{3pt}{$\hbox to 28pt{\hfill${\scriptstyle 2.4}$\hfill}$}
\ins{0pt}{21pt}{$\hbox to 20pt{\hfill${\scriptstyle 0.0}$}$}
\ins{0pt}{50pt}{$\hbox to 20pt{\hfill${\scriptstyle 0.2}$}$}
\ins{0pt}{79pt}{$\hbox to 20pt{\hfill${\scriptstyle 0.4}$}$}
\ins{0pt}{108pt}{$\hbox to 20pt{\hfill${\scriptstyle 0.6}$}$}
\ins{0pt}{137pt}{$\hbox to 20pt{\hfill${\scriptstyle 0.8}$}$}
\ins{0pt}{166pt}{$\hbox to 20pt{\hfill${\scriptstyle 1.0}$}$}
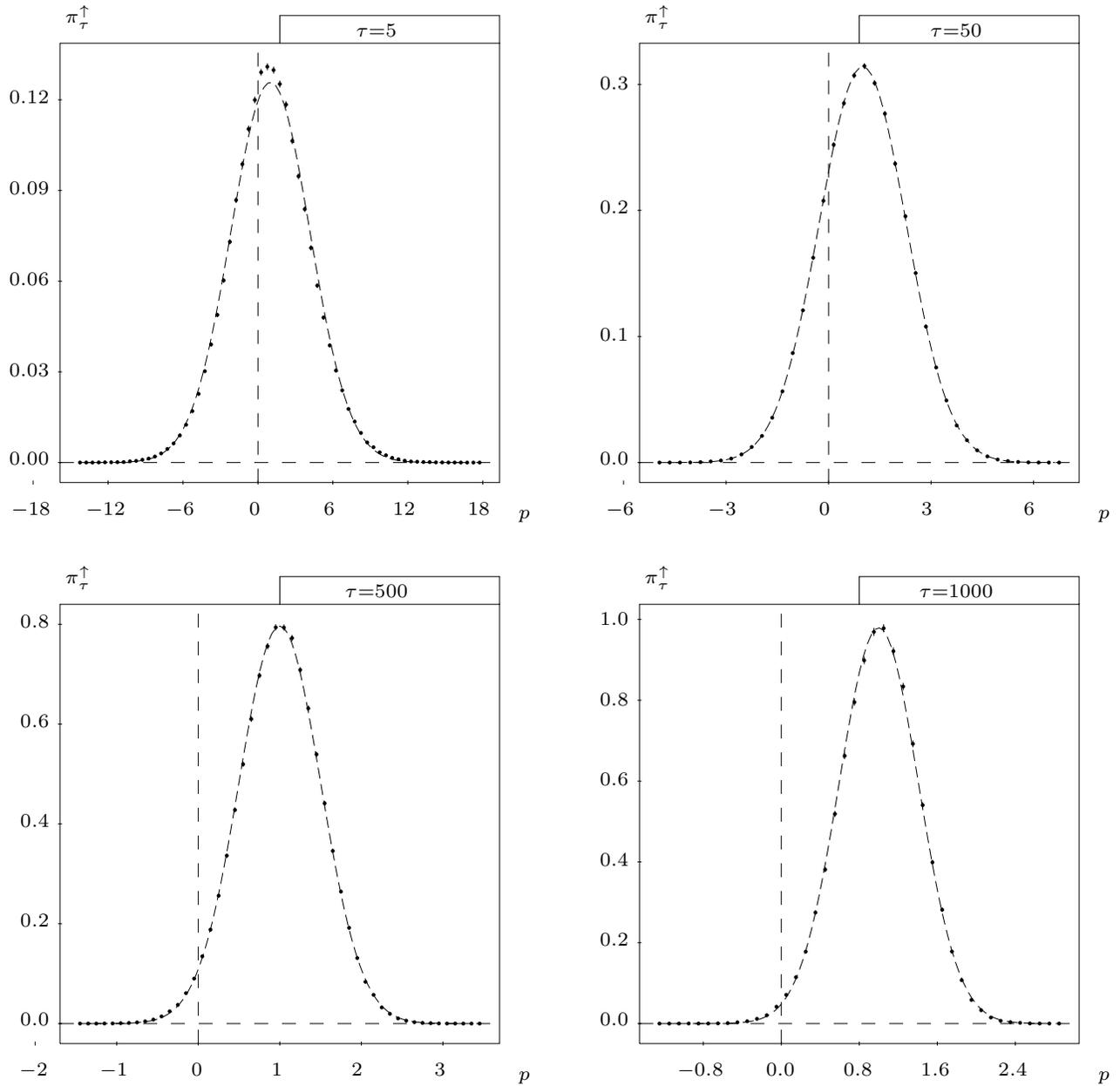
}{Fig. \equ(fig10): comparison between the distribution
$\pi_\tau^\su(p)$ and a Gaussian with the same variance and mean. The
errorbar are smaller than the dimension of the points.}{fig10}

As in the previous section we can assume that the pair $(\sigma^\su,
\sigma^\giu)$ represents a bidimensial Gaussian variable with mean
$(1,1)$ and covariance matrix:

$${\cal C}_\tau=\pmatrix{C_\tau^\su&C_\tau^{\su\giu}\cr
C_\tau^{\su\giu}&C_\tau^{\giu}\cr}\Eq(cm)$$
where

$$C^\su_\tau=\langle(\sigma^\su_\tau)^2\rangle-1\qquad
C^\giu_\tau=\langle(\sigma^\giu_\tau)^2\rangle-1\Eq(cmi)$$

Form now on we will suppose that $C_\tau^\su=C_\tau^{\giu}$. This is
well verified in the experiment and seems natural from the symmetry
properties of the system (see however footnote 1).  With the above
Gaussian hypothesis we can again try to follow the behavior of
$\chi^\su_\tau$ for large value of $\tau$ using the expression:

$$\chi^\su_\tau=\frac{2}{\langle \alpha^\su\rangle \tau C^\su_\tau}\Eq(chi_s)$$
Fig. \equ(fig11) shows the results of this evaluation.

\dimen2=25pt \advance\dimen2 by 126pt
\dimen3=204pt
\dimenfor{\dimen3}{166pt}
\eqfigfor{fig_fffs}{
\ins{184pt}{0pt}{$\hbox to20pt{\hfill ${\scriptstyle \tau}$\hfill}$}
\ins{12pt}{\dimen2}{$\hbox to25pt{\hfill ${\scriptstyle \chi^\su_\tau}$\hfill}$}
\ins{103pt}{144pt}{$\hbox to83pt{\hfill ${\scriptstyle N=10}$\hfill}$}
\ins{6pt}{3pt}{$\hbox to 37pt{\hfill${\scriptstyle 0}$\hfill}$}
\ins{44pt}{3pt}{$\hbox to 37pt{\hfill${\scriptstyle 10000}$\hfill}$}
\ins{82pt}{3pt}{$\hbox to 37pt{\hfill${\scriptstyle 20000}$\hfill}$}
\ins{120pt}{3pt}{$\hbox to 37pt{\hfill${\scriptstyle 30000}$\hfill}$}
\ins{158pt}{3pt}{$\hbox to 37pt{\hfill${\scriptstyle 40000}$\hfill}$}
\ins{0pt}{21pt}{$\hbox to 14pt{\hfill${\scriptstyle 0.1}$}$}
\ins{0pt}{41pt}{$\hbox to 14pt{\hfill${\scriptstyle 0.3}$}$}
\ins{0pt}{62pt}{$\hbox to 14pt{\hfill${\scriptstyle 0.5}$}$}
\ins{0pt}{82pt}{$\hbox to 14pt{\hfill${\scriptstyle 0.7}$}$}
\ins{0pt}{102pt}{$\hbox to 14pt{\hfill${\scriptstyle 0.9}$}$}
\ins{0pt}{122pt}{$\hbox to 14pt{\hfill${\scriptstyle 1.1}$}$}
}{\ins{184pt}{0pt}{$\hbox to20pt{\hfill ${\scriptstyle \tau}$\hfill}$}
\ins{12pt}{\dimen2}{$\hbox to25pt{\hfill ${\scriptstyle \chi^\su_\tau}$\hfill}$}
\ins{103pt}{144pt}{$\hbox to83pt{\hfill ${\scriptstyle N=20}$\hfill}$}
\ins{6pt}{3pt}{$\hbox to 37pt{\hfill${\scriptstyle 0}$\hfill}$}
\ins{44pt}{3pt}{$\hbox to 37pt{\hfill${\scriptstyle 10000}$\hfill}$}
\ins{82pt}{3pt}{$\hbox to 37pt{\hfill${\scriptstyle 20000}$\hfill}$}
\ins{120pt}{3pt}{$\hbox to 37pt{\hfill${\scriptstyle 30000}$\hfill}$}
\ins{158pt}{3pt}{$\hbox to 37pt{\hfill${\scriptstyle 40000}$\hfill}$}
\ins{0pt}{21pt}{$\hbox to 14pt{\hfill${\scriptstyle 0.2}$}$}
\ins{0pt}{39pt}{$\hbox to 14pt{\hfill${\scriptstyle 0.4}$}$}
\ins{0pt}{57pt}{$\hbox to 14pt{\hfill${\scriptstyle 0.6}$}$}
\ins{0pt}{75pt}{$\hbox to 14pt{\hfill${\scriptstyle 0.8}$}$}
\ins{0pt}{93pt}{$\hbox to 14pt{\hfill${\scriptstyle 1.0}$}$}
\ins{0pt}{112pt}{$\hbox to 14pt{\hfill${\scriptstyle 1.2}$}$}
\ins{0pt}{130pt}{$\hbox to 14pt{\hfill${\scriptstyle 1.4}$}$}
}{\ins{184pt}{0pt}{$\hbox to20pt{\hfill ${\scriptstyle \tau}$\hfill}$}
\ins{12pt}{\dimen2}{$\hbox to25pt{\hfill ${\scriptstyle \chi^\su_\tau}$\hfill}$}
\ins{103pt}{144pt}{$\hbox to83pt{\hfill ${\scriptstyle N=30}$\hfill}$}
\ins{6pt}{3pt}{$\hbox to 37pt{\hfill${\scriptstyle 0}$\hfill}$}
\ins{44pt}{3pt}{$\hbox to 37pt{\hfill${\scriptstyle 10000}$\hfill}$}
\ins{82pt}{3pt}{$\hbox to 37pt{\hfill${\scriptstyle 20000}$\hfill}$}
\ins{120pt}{3pt}{$\hbox to 37pt{\hfill${\scriptstyle 30000}$\hfill}$}
\ins{158pt}{3pt}{$\hbox to 37pt{\hfill${\scriptstyle 40000}$\hfill}$}
\ins{0pt}{19pt}{$\hbox to 14pt{\hfill${\scriptstyle 0.3}$}$}
\ins{0pt}{36pt}{$\hbox to 14pt{\hfill${\scriptstyle 0.5}$}$}
\ins{0pt}{53pt}{$\hbox to 14pt{\hfill${\scriptstyle 0.7}$}$}
\ins{0pt}{70pt}{$\hbox to 14pt{\hfill${\scriptstyle 0.9}$}$}
\ins{0pt}{87pt}{$\hbox to 14pt{\hfill${\scriptstyle 1.1}$}$}
\ins{0pt}{105pt}{$\hbox to 14pt{\hfill${\scriptstyle 1.3}$}$}
\ins{0pt}{122pt}{$\hbox to 14pt{\hfill${\scriptstyle 1.5}$}$}
\ins{0pt}{139pt}{$\hbox to 14pt{\hfill${\scriptstyle 1.7}$}$}
}{\ins{184pt}{0pt}{$\hbox to20pt{\hfill ${\scriptstyle \tau}$\hfill}$}
\ins{12pt}{\dimen2}{$\hbox to25pt{\hfill ${\scriptstyle \chi^\su_\tau}$\hfill}$}
\ins{103pt}{144pt}{$\hbox to83pt{\hfill ${\scriptstyle N=40}$\hfill}$}
\ins{6pt}{3pt}{$\hbox to 37pt{\hfill${\scriptstyle 0}$\hfill}$}
\ins{44pt}{3pt}{$\hbox to 37pt{\hfill${\scriptstyle 10000}$\hfill}$}
\ins{82pt}{3pt}{$\hbox to 37pt{\hfill${\scriptstyle 20000}$\hfill}$}
\ins{120pt}{3pt}{$\hbox to 37pt{\hfill${\scriptstyle 30000}$\hfill}$}
\ins{158pt}{3pt}{$\hbox to 37pt{\hfill${\scriptstyle 40000}$\hfill}$}
\ins{0pt}{17pt}{$\hbox to 14pt{\hfill${\scriptstyle 0.4}$}$}
\ins{0pt}{34pt}{$\hbox to 14pt{\hfill${\scriptstyle 0.6}$}$}
\ins{0pt}{51pt}{$\hbox to 14pt{\hfill${\scriptstyle 0.8}$}$}
\ins{0pt}{68pt}{$\hbox to 14pt{\hfill${\scriptstyle 1.0}$}$}
\ins{0pt}{86pt}{$\hbox to 14pt{\hfill${\scriptstyle 1.2}$}$}
\ins{0pt}{103pt}{$\hbox to 14pt{\hfill${\scriptstyle 1.4}$}$}
\ins{0pt}{120pt}{$\hbox to 14pt{\hfill${\scriptstyle 1.6}$}$}
\ins{0pt}{137pt}{$\hbox to 14pt{\hfill${\scriptstyle 1.8}$}$}
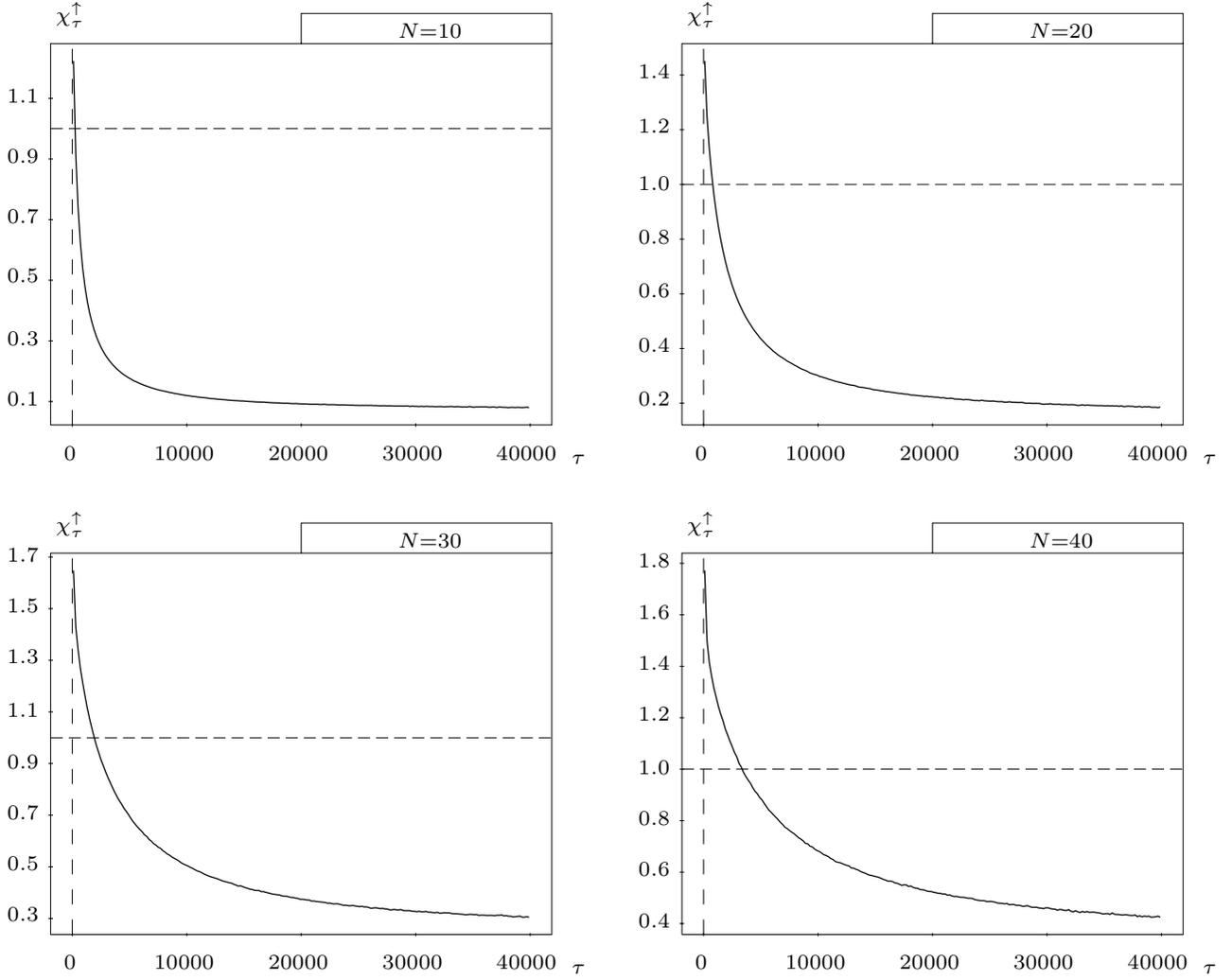
}{Fig. \equ(fig11): Behavior of $\chi^\su_\tau$ as a function of $\tau$ for 
$N=10,20,30,40$.}{fig11}

Finally we observe that we can write:

$$C^{\su\giu}_\tau=2C_\tau-C^\su_\tau\Eq(som)$$

Hence, if we assume gaussianity and take $\tau$ large enough that we
can consider $\chi_\tau=1$, gives

$$K_\tau=\chi^\su_\tau-1=\frac{C^{\su\giu}_\tau}{C^\su_\tau}\Eq(K)$$

We can conclude this discussion by observing that, for finite $N$, the
fluctuation law seems to be invalid if we take in consideration only
half of the entropy production. Nevertheless it is interesting to note
that $x^{\su,\giu}_\tau(p)$ still look linear and that the slope
$\chi^{\su,\giu}_\tau$ seems to reach a finite limit as $\tau$ goes to
infinity. Moreover this limit is very small for small $N$ and seems to
increase with $N$. It would be intersting to fit the curve in
Fig. \equ(fig11) to give an estimate of the limiting value of
$\chi_\tau^\su$ as $\tau\to\infty$.  We observe that, as for
$\chi_\tau$, we can relate the approach to a limiting value of
$\chi^\su_\tau$ with the decay properties of
$D^\su(t)=\langle\sigma^\su(S^t(\cdot))\sigma^\su(\cdot)\rangle -1$.
In fact an equation identical to \equ(dec) holds for $C^\su_\tau$ with
$D^\su(t)$ in the place of $D(t)$. A comparison of Fig. \equ(fig6) and
Fig. \equ(fig11) immediately shows that the decay of correlations of
$\sigma^\su$ is much slower that that of $\sigma$. 
If we assume that $D^\su(t)\simeq t^{-\beta}$, with $\beta>0$, we get that:

$$\chi_\tau^\su=\chi^\su_\infty+\frac{{\chi^\su}^{\prime}}{\tau}+
\frac{{\chi^\su}^{\prime\prime}}{\tau^{\beta-1}}$$

If $\beta$ is smaller than 2 the third term dominate, in the
asymptotic behavior on the second one. 
One can try to fit $\beta$ looking at the log-log plot of
$\chi^\su_\tau$ and then using a least square fit to find the other
coefficient. Unfortunately the log-log plot does not give a precise
answer, so we can only say that $\beta$ is probably less than
$2$. This implies that the approach to the limit of
$\chi^{\su,\giu}_\tau$ is very slow and its limiting value change
considerably depending which $\beta$ one uses. At the end we are unable
to give a quantitative estimate of $\chi^{\su,\giu}_\infty$ and we can
just say that it is non zero and increases with $N$.

\newsec{Conclusion}

Our numerical results show that the Fluctuation Theorem, eq.\equ(stau),
is well verified if we consider the total phase space contraction
$\sigma$ of the model described in sec. 1 sub {\it A}. For $\tau$ small
enough we were able to construct the distribution $\pi_\tau(p)$ and
check directly the validity of eq. \equ(stau). For higher $\tau$ we
used a Gaussian hypothesis to compute the slope $\chi_\tau$ of
$x_\tau(p)$ obtaining again a very good agrrement with the theoretical
prediction. We observe that if we interpret the Gaussianity of
$\pi_\tau(p)$ as a Central Limit Theorem effect we get that
$x_\tau(p)$ is linear but no conclusion con be drawn on the slope
$\chi_\tau$. The validity of the fluctuation theorem predictions,
toghether with the gaussian hypothesis, imply a Green-Kubo formula out
of equilibrium as can be seen by combining eq. \equ(chi) and
eq. \equ(dec) (see also \cita{BGG}). In this sense we can think that 
our results depend in part on the fact that we are not very far from
equilibrium. Increasing the shear make the experiment much harder
becouse the probability of observing negative fluctuations decreases
very rapidly with the shear. Moreover we expect that the gaussianity
of $\pi_\tau(p)$ will be destroyed. We think, however, that if the
shear is not too big, \ie if the attractor can still be considered
dense in phase space (see \cita{BG}), then the fluctuation theorem will
holds (see \cita{LLP} for a situation in wich $\pi_\tau(p)$ is clearly
not Gaussian).

We also checked the validity of the fluctuation theorem for the
partial phase space contraction $\sigma^\su$ and $\sigma^\giu$. In
this situation we see that $\pi^{\su,\giu}$ is still well aproximated
by a Gaussian so that $x^\su_\tau(p)$ appears to be linear. The slope
$\chi_\tau^\su$ however behave in a different way from what one would
expect. Two interesting features of this behaviour are:

\item{1)} It saturates very slowly. The fact that we considered, in the
definition of $\sigma^{\su,\giu}_\tau$ a fixed number of collisions
$\tau$ with both walls clearly create a negative correlation between
$\sigma^{\su}_\tau$ and $\sigma^{\giu}_\tau$. This correlation can be
roughly estimated as $\tau^{-1}$ so that it probably cannot account
for the slow behavior found in sec. 3. That behavior shows a strong
correlation between the two wall probably due to the slow decay of
fluctuation the density near the walls. We do not have a clear
understanding of this phenomenon and we hope to come back on it in a
forthcaming work.

\item{2)} The limiting value $\chi_\infty^{\su,\giu}$ in not easy to
estimate but is clearly smaller than 1. Although it seems to increase
with the number of particle we do not have enough data to draw any
conclusion on its behaviour with $N$. Nevertheless the existence of a
limiting non-zero value suggests the validity of a ``local fluctuation
theorem'' in wich the slope 1 is replaced by some other value. This
behavior may have relation to the recent experimental result on the
local entropy production (supposed to be the equivalent of a phase
space contraction) in a fluid moving in a convective cell \cita{CLR}.
They find there a clear linear behaviour but the slope seems to be
different from one. The rusults are, however, still unclear and no real
comparison can be made.
\*
\0{\bf Aknowledgemets}
F.B. is endebted to E.G.D. Cohen and G. Gallavotti for many helpuful
discussion and suggestion. N.I.C. was partly supported by NFS grant
DMS-9654622. F.B and J.L.L were in part supported by NFS Grant DMR-9523266.

\rife{ECM2}{ECM2}{D.J. Evans, E.G.D. Cohen, G.P. Morris, ``Probability
of Second Law Violations in Shearing Steady Folws'',
Phys. Rew. Letters {\bf 71}, 2401--2404 (1993).}
\rife{CL}{CL}{N.I. Chernov, J.L. Lebowitz, ``Stationary Nonequilibrium
States for Boundary Driven Hamiltonian Systems'' JSP {\bf 86},
953--990 (1997).}
\rife{PD}{DP}{C. Dellago, H.A. Posch, ``Lyapinov Instability of the
Boundary Driven Chernov--Lebowitz Model for Stationary Shear Flow''
JSP {\bf 88}, 825--842 (1997).}
\rife{GC}{GC}{G. Gallavotti, E.G.D. Cohen, ``Dynamical ensamble in a
stationary states'' Jour. Stat. Phys {\bf 80}, 931--970 (1995).}
\rife{BGG}{BGG}{F. Bonetto, G. Gallavotti, P.L. Garrido, ``Chaotic
principle: an experimentat test'', Physica D {\bf 105}, 226--252 (1997).}
\rife{G}{G}{G. Gallavotti, ``Extension of the Onsager reciprocity to
large field and the chaotic hypothesis'' Phys. Rev. Lett. {\bf 77},
4334--4337 (1996).}
\rife{AAVV}{AAVV}{G. Gallavotti, ``New Methods in Nonequilibrium
Gases and Fluids'', archived in mp-arc at www.ma.utexas.edu \#96-533 and
D. Ruelle, ``Dynamical Systems Approach to Nonequilibrium
Statistical Mechanics: An Introduction'', IHES/Rutgers, Lecture Notes, 1997. }
\rife{BB}{BB}{D.J. Evans, G.P. Morris, {\it Statistical Mechanics of
Nonequilibrium Liquids}, Accademic Press (1990).}
\rife{ST}{ST}{G. Eyink, J.L. Lebowitz, H. Sphon, ``Lattice Gas Model
in Contact with Stochastic Reservoires: Local Equilibrium and
Relaxation to Steady States'' Comm. Math. Phys. {\bf 140}, 119--132 (1991).}
\rife{EE}{EE}{Z. Cheng, P.L. Garrido, J.L. Lebowitz, ``Correlations in
Stationary Nonequilibrium Systems with Conservative Anisotropic
Dynamics'' Europhysics Lett. {\bf 14}, 507--513 (1991).}
\rife{BG}{BG}{F.Bonetto, G. Gallavotti, ``Chaotic principle, reversibility
and coarse graining'', Comm. Math. Phys {\bf 189}, 263--276 (1997).}
\rife{LLP}{LLP}{S. Lepri, R. Livi, A. Politi, ``Energy transport in
anharmonic lattice close and far from equilibrium'', preprint,
archived in xxx.lanl.gov cond-mat \#9709195.}
\rife{CLR}{CLR}{S. Ciliberto,``An experimental verification of the
Gallavotti-Cohen fluctuation theorem'', preprint.}
\biblio

\fine{}

\ciao